\documentclass[10pt]{article}

\usepackage{amsmath}
\usepackage{amssymb}

\usepackage{graphicx}

\usepackage{cite}

\usepackage{rotating}
\usepackage{amsfonts}

\usepackage{color} 

\newcommand{\deltat}{\ensuremath{\Delta t}}

\usepackage[labelfont=bf,labelsep=period,justification=raggedright]{caption}

\makeatletter
\renewcommand{\@biblabel}[1]{\quad#1.}
\makeatother

\date{}

\begin{document}

\begin{flushleft}
{\Large
\textbf{The developmental dynamics of terrorist organizations}
}
\\
Aaron Clauset$^{1,2,3,\ast}$ and Kristian Skrede Gleditsch$^{4,5,\ast}$
\\
\bf{1} Department of Computer Science, University of Colorado, Boulder, CO, USA
\\
\bf{2} {BioFrontiers} Institute, University of Colorado, Boulder, CO, USA
\\
\bf{3} Santa Fe Institute, Santa Fe, NM, USA
\\
\bf{4} Department of Government, University of Essex, Wivenhoe Park, Colchester, UK
\\
\bf{5} Centre for the Study of Civil War, Oslo, Norway
\\
$\ast$ E-mail: Corresponding aaron.clauset@colorado.edu
\end{flushleft}

\section*{Abstract}
We identify robust statistical patterns in the frequency and severity of violent attacks by terrorist organizations as they grow and age. Using group-level static and dynamic analyses of terrorist events worldwide from 1968--2008 and a simulation model of organizational dynamics, we show that the production of violent events tends to accelerate with increasing size and experience. This coupling of frequency, experience and size arises from a fundamental positive feedback loop in which attacks lead to growth which leads to increased production of new attacks. In contrast, event severity is independent of both size and experience. Thus larger, more experienced organizations are more deadly because they attack more frequently, not because their attacks are more deadly, and large events are equally likely to come from large and small organizations. These results hold across political ideologies and time, suggesting that the frequency and severity of terrorism may be constrained by fundamental processes.


\section*{Introduction}
Much research on patterns in terrorism has been inspired by particular historic events and ``waves'' of specific forms of terrorist attacks~\cite{rapoport:2004,rosenfeld:2011}. Just as the rise in international skyjackings in the 1970s led to a resurgence of studies of terrorism, the 11 September 2001 attacks renewed interest in why groups resort to terrorism, the specific choice of attack targets, and the relative effectiveness of particular counterterrorism measures. As a result, many researchers have developed typologies of specific forms of terrorism and highlighted the distinctiveness of different terrorist groups. By contrast, in this manuscript we examine whether there are fundamental patterns in the frequency and severity (number of deaths) of deadly events carried out by terrorist organizations and what mechanisms might generate them.

Little research on terrorism has focused on directly modeling individual event frequency and severity, and the way these change over an organization's lifetime. When deaths are considered, they are typically aggregated and used as a covariate to understand other aspects of terrorism, e.g., trends over time~\cite{enders:sandler:2000,enders:sandler:2002}, the when, where, what, how and why of the resort to terrorism~\cite{brown:dalton:hoyle:2004,enders:sandler:2006,valenzuela:etal:2010}, differences between organizations~\cite{asal:rethemeyer:2008}, or the incident rates or outcomes of events~\cite{enders:sandler:2000,enders:etal:2011}. Such efforts have used time series analysis~\cite{enders:sandler:2000,enders:sandler:2002,enders:etal:2011}, qualitative models or human expertise of specific scenarios, actors, targets or attacks~\cite{wulf:etal:2003} or quantitative models based on factor analysis~\cite{li:2005,pape:2003}, social networks~\cite{sageman:2004,desmarais:cranmer:2011} or formal adversarial interactions~\cite{major:1993,kardes:hall:2005,enders:sandler:2006}.

Our approach is different and complementary to these approaches, focusing on global trends and patterns in the frequency and severity of events~\cite{richardson:1960,cederman:2003,lacina:2006,clauset:etal:2007,mcmorrow:2009,clauset:gleditsch:2009,bohorquez:etal:2009,clauset:etal:2010:b,johnson:etal:2011}, rather than on event particulars or motivations. 
By focusing our analysis at the global scale, the importance of individual decisions in specific contexts is in fact lessened, due to the central limit theorem and the rough independence of individual events; as a result, the importance of generic non-strategic processes is enhanced and these processes, if any, may be studied. Explanations of such patterns must thus focus on processes or constraints that are independent of variations in context or specific motivation and may include physical constraints, network effects and endogenous population dynamics, which are well suited to explain the behavior of strategically unc\"oordinated populations of actors~\cite{clauset:etal:2010:b}. This approach to investigating the fundamental laws of terrorism has much in common with that of statistical physics, in which the self-averaging properties of independent events allows for interesting population-level properties to emerge from microscopic system chaos. This statistical physics-style approach is increasingly being applied to study complex social systems~\cite{holme:ghoshal:2006,gutfraind:2010,turchin:2012}, yielding a number of novel insights.

Here, we aim to shed new light on the fundamental processes governing the frequency and severity of terrorist events by studying their statistical relationship with the organizations that generate them. Our aim is to identify global patterns in these relationships and to explain their origin mechanistically. We employ a combination of disaggregated data analysis, studying a large database of terrorist events worldwide from 1968--2007, statistical modeling and inference, computational modeling and regression analysis to validate our mechanistic hypotheses. By shedding new light on these large-scale patterns and trends in terrorism, and on how such patterns emerge from local-level behaviors, this large-scale statistical or pattern-based approach can supplement formal models of strategic interactions, inform counter-terrorism policy and clarify our general ability to forecast or anticipate future terrorist events or trends.

\subsection*{Patterns in global conflict}

A pattern-based approach to studying conflict owes much to the seminal work in the early 20th century of Lewis Fry Richardson---a physicist and meteorologist known for collecting data on conflicts (``deadly quarrels''), modeling arms races using differential equations, as well as early contributions to understanding the frequencies and severities of wars. Specifically, Richardson~\cite{richardson:1941,richardson:1948} identified the remarkable pattern that the frequency of wars decays like the inverse power of their severity.
(Power-law distributions can indicate unusual underlying or endogenous processes, e.g., feedback loops, network effects, self-organization or optimization. From a purely statistical perspective, power-law distributions generate large events orders of magnitude more often than we would expect under a Normal assumption. Recently, power-law distributions have been identified in a wide range of social and biological systems~\cite{clauset:etal:2009}. See~\cite{kleiber:kotz:2003},~\cite{mitzenmacher:2004} and~\cite{newman:2005} for reviews, or Appendix~A of~\cite{clauset:wiegel:2010} for a gentle introduction.)
This empirical pattern implies that there is no fundamental statistical difference between rare but catastrophic wars and more common but less severe wars---the likelihoods of both are described by a single mathematical function:
\begin{align}
\Pr({\rm event~with~severity}~x) \propto x^{-\alpha} \enspace , \nonumber
\end{align}
where $x$ counts the number of fatalities (severity) and $\alpha$ is the ``scaling exponent,'' which controls how quickly the frequency decreases as severity increases. It also implies that the underlying social and political processes for both large and small wars may be fundamentally the same, i.e., large wars may simply be ``scaled up'' versions of small wars. In general, the identification of a power law implies that studying the statistically more common events can shed light on certain aspects of extremely rare events.
(Seismologists study large earthquakes in this way: the frequencies of both large and small quakes follow a power-law distribution, called the Gutenberg-Richter Law, and the physical processes that generate both small and large quakes are fundamentally the same.)

Recently, Clauset et al.~\cite{clauset:etal:2007,clauset:etal:2009} showed that this same pattern---a power-law, ``Richardson's Law''---also holds for the frequency of severe terrorist attacks (reported fatalities) worldwide, while~\cite{bohorquez:etal:2009} suggest a similar pattern for events within insurgencies. The power-law pattern in terrorism is highly robust: it persists over the past 40 years despite large structural and political changes in the international system and is independent of the type of weapon used (explosives, firearms, arson, knives, etc.), the emergence and increasing popularity of suicide attacks, the demise of many individual terrorist organizations, and the economic development of the target country.

Thus, fundamental regularities in terrorism can and do emerge at
the global level despite the highly contingent and
context-specific nature of the individual attacks, conflicts and
decisions. Insights into how these patterns' arise will likely
shed new light on the underlying social or political processes
that drive and constrain global trends and on effective policies
for responding to or managing those processes.

\section*{Methods}
We consider the frequency and severity of attacks over the lifetime of individual terrorist organizations, and the question of whether organizations exhibit common statistical patterns in these behaviors. We argue that organization size (number of personnel) plays a fundamental role in limiting the overall frequency, but not the severity, of violent events by a group. The key idea is that organization size and its overall production rate of events are linked. If events lead to growth in any way, then this link implies a positive feedback loop in which each attack increases the production rate of future attacks. Thus, a terrorist organization can be viewed as a kind of factory whose principal product is political violence, and whose proceeds are reinvested in increased production capacity.

To test these ``developmental dynamics'' hypotheses, we present novel statistical analyses of the behavior of nearly 400 terrorist organizations worldwide over the period 1968--2008. We find strong evidence for precisely this kind of generic acceleration in event production. This supports the notion that an organization's available labor, i.e., the size of its militant wing, is a fundamental constraint on the overall frequency of its attacks. We further show that the rate at which an organization cycles through the positive feedback loop can depend on covariates like its political ideology, with religiously-motivated organizations accelerating (growing) the fastest. In contrast, we find no evidence that event severity depends on organizational size or experience. Instead, the distribution of attack severities follows a rough form of Richardson's Law independent of size, experience or political motivation.

These results imply that very large events are equally likely to be generated by small groups as by large groups, and that larger organizations are indeed more deadly~\cite{asal:rethemeyer:2008}, not because their individual attacks are systematically more spectacular but because they typically carry out many more attacks. That is, the size of the beast directly determines the overall level of terror activity (frequency) but not the quality (severity) of those actions.

Recently, Johnson et al.~\cite{johnson:etal:2011} used a similar approach to analyze the timing of events in the Iraq and Afghanistan conflicts, which was in turn based on an earlier version of this manuscript~\cite{clauset:gleditsch:2009}. Although similar statistical patterns to the ones we describe here were observed in those conflicts, a different explanation was offered for their origin. We will revisit this comparison and comment on the problems our statistical results pose for the explanation offered by~\cite{johnson:etal:2011}.

\subsubsection*{Impact of Size on Frequency}
\begin{enumerate}
\item[H1.] {\em Labor-constraints}: the overall production rate of violent events by an organization depends on its size, and thus the time between consecutive attacks $\Delta t$ is roughly inversely proportional to the size $s$ of the organization. Mathematically, $s\propto 1/\Delta t$.
\end{enumerate}
In other words, the production of terrorist events cannot be automated. If this were possible, organizations could produce arbitrary numbers of events without needing to grow in size, much like a fully automated factory requires essentially no human personnel to function.
(In this light, {\em cyber terrorism} is an interesting case: it remains unclear to what degree the planning and execution of cyber terrorist attacks can be done automatically, by computers. Our current belief is that cyber terrorism is also not mass produceable and thus some labor constraint will persist, although it may be substantially lessened relative to physical terrorism.)
Instead, we argue that each terrorist event requires significant human involvement, e.g., to conceive, plan and execute it. This requirement for human effort implies that for the production rate of an organization to decrease, it must add additional members to produce them. And, the resultant increased rate occurs not because more hands make any individual event proceed more quickly, but because multiple events may be carried out in parallel. That is, the overall production rate of the organization is like the production rate of an entire factory; as the factory (organization) adds internal independent production lines (terrorist cells), the effective time between new events falls even though each production line operates at a constant rate.

It is important to recognize that H1 does not imply that the only way to increase the group-level production rate of attacks is through organizational growth. Indeed, many aspects of event production surely do benefit from technology or efficiency improvements~\cite{dutton:thomas:1984,argote:1993,jackson:etal:2005,thompson:2010}. Instead, H1 implies that such factors can only moderate,  not eliminate, the fundamental constraint that size places on production. To the extent that these factors decrease the time between an organization's events, the literature on learning suggests that the overall impact will be modest~\cite{thompson:2010}. In contrast, increases in labor, which allow many terrorist cells to operate in parallel, can lead to much larger improvements.

Finally, we note that this constraint should be strongest for small organizations, who likely have the worst access to efficiency-improving resources like specialized personnel, training facilities or factories and who may reap the largest benefit, e.g., media visibility, from striving to maximize their event production. Because most organizations begin small and grow over time, this should be most evidence early in the lifetime of an organization.
(A spatial corollary of H1 is that if an ``organization'' is defined as those militants within some geographic locale, e.g., a province or district, then the frequency of events within that locale will be roughly inversely proportional to the number of militants there. That is, the $s\propto 1/\Delta t$ relationship should hold when both $s$ and $\Delta t$ are defined by a geographic boundary. Organizational ``growth'' can then be understood as either immigration or recruitment  of new militants.)

\subsubsection*{Events, Recruitment and Growth}
What role do attacks play in changing organizational size? If an event gains the organization wider visibility among potential members or sympathizers, the organization may grow in size as a result of that event. (Decreases in size are likely driven by distinct social processes (see~\cite{cronin:2009}), which we do not consider here.)
\begin{enumerate}
\item[H2.] {\em Event-recruitment}: organizational growth (increased $s$) is partly driven by recruitment associated with the production of new events (increased $k$), i.e., events lead to recruitment which leads to organizational growth. Mathematically, ${\rm d}s/{\rm d}k>0$.
\end{enumerate}
H2 does not imply that growth comes only from violence-related recruitment. So long as recruitment is partly based on the production of violent events, H2 implies a correlation between increases in size and increased event production

\subsubsection*{Frequency Acceleration}
Together, H1 and H2 imply a positive feedback loop in which attacks lead to recruitment which leads to organizational growth and thus an increased group-level production of new attacks. So long as a portion of the growth is allocated to producing additional events, i.e., so long as the militant wing grows with the overall organization, H1 and H2 jointly imply H3.
\begin{enumerate}
\item[H3.] {\em Frequency-acceleration}: as an organization carries out more attacks (increased $k$), the time between subsequent attacks $\Delta t$ decreases. Mathematically, ${\rm d}\Delta t/{\rm d}k<0$.
\end{enumerate}
That is, H1 predicts $s\propto 1/\Delta t$ while H2 predicts ${\rm d}s/{\rm d}k>0$. Eliminating the common factor of $s$ yields the prediction that ${\rm d}\Delta t/{\rm d}k<0$, in which the continued production of violent events produces a decreasing delay between those events.
(This dynamical relationship produces a similar pattern to that observed in ``learning'' or ``progress curves,'' in which continued production covaries with lowered production costs or time~\cite{dutton:thomas:1984,argote:etal:1995,thompson:2010}. Although the pattern is similar, the mechanism is different.)

\subsubsection*{Impact of Size on Severity}
Increased size may bring greater access to capital and skilled labor, e.g., experienced professionals, advanced arms, intelligence, etc., and thus more spectacular attacks.
\begin{enumerate}
\item[H4.] {\em Severity-increase}: the severity~$x$ of a new attack increases with organizational size $s$ and, via H2, the number of attacks $k$. Mathematically, ${\rm d}x/{\rm d}s>0$ and ${\rm d}x/{\rm d}k>0$, respectively.
\end{enumerate}
Combined with H2, H3 implies that attacks by experienced, larger groups should be consistently and significantly more deadly than those of less experienced or smaller groups.

H4 assumes a tangible benefit for maximizing the severity of attacks, e.g., to gain wider visibility for the organization's cause or to demonstrate power or resolve. Such incentives are not foregone conclusions: severe attacks may also attract harsh attention from state-level actors, leading to repression, police action or the destruction of physical or financial resources. They may also induce counter-productive effects on potential sympathizers, e.g., due to the shockingness of spectacular events. As a result, we consider the theoretical argument supporting the severity-increase hypothesis to be marginal.

\section*{Results}
\subsection*{Model of terrorist organizations}
\begin{figure}[t]
\begin{center}
\includegraphics[scale=0.65]{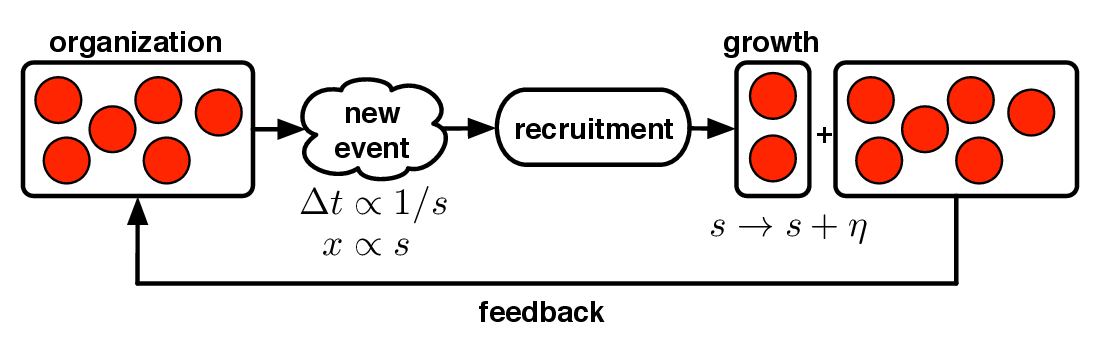}
\end{center}
\caption{\textbf{A model of terrorist organizations.} A schematic illustrating the feedback loop relationship between size $s$ and the frequency and severity of attacks: the delay between subsequent attacks $\Delta t$ is inversely related to an organization's size $s$ while the severity of subsequent attacks $x$ grows with $s$; new events lead to recruitment which leads to growth, which increases the size variable $s$.}
\label{fig:schematic}
\end{figure}

To illustrate these interactions between an organization's size and the frequency and severity of attacks over its lifetime, we construct a simple model of a terrorist organization's development (see Figure~\ref{fig:schematic} for a schematic).

Historically, terrorist organizations begin as a small collections of terrorism-inclined individuals~\cite{hoffman:1998}. Let this initial collection be composed of roughly $\eta$ individuals, which denotes the typical or characteristic size of a terrorist cell. The particular value of $\eta$ is not important, but may depend political ideology, socio-economic context~\cite{krueger:2007}, the attack's target, etc. The cell plans and conducts its first attack, which gains it some visibility, via either traditional media coverage or informal channels. Subsequent recruitment yields a number of additional members~$\nu$~(H2), and now the organization is larger. Again, the particular value of~$\nu$ is not important, but likely depends on context-specific factors.

Each cell continues planning and carrying out new attacks, roughly once every $\tau$ days (H1). Newly recruited members form new cells, of size $\eta$~(H1) and new cells plan and carry out their own attacks in parallel. It is this parallelism that allows the larger organization to appear to be acting more quickly, even though the planning time $\tau$ for any particular event remains fixed. An attack by any cell leads to overall organizational growth via recruitment~(H2), which in turn increases the organization's overall production rate of attacks by adding new cells~(H3). Finally, as the group grows, the increased manpower also increases its ability to carry out more severe events~(H4), e.g., because more supporting roles allow better surveillance, access to better equipment, etc.

Coordinating the activities of these additional individuals, or the development of non-violent initiatives like a political wing or the provision of social services, will draw some members away from these militant activities. However, so long as recruitment continues to grow the number of militant cells, the positive feedback loop remains.

\begin{figure}[t]
\begin{center}
\includegraphics[scale=0.435]{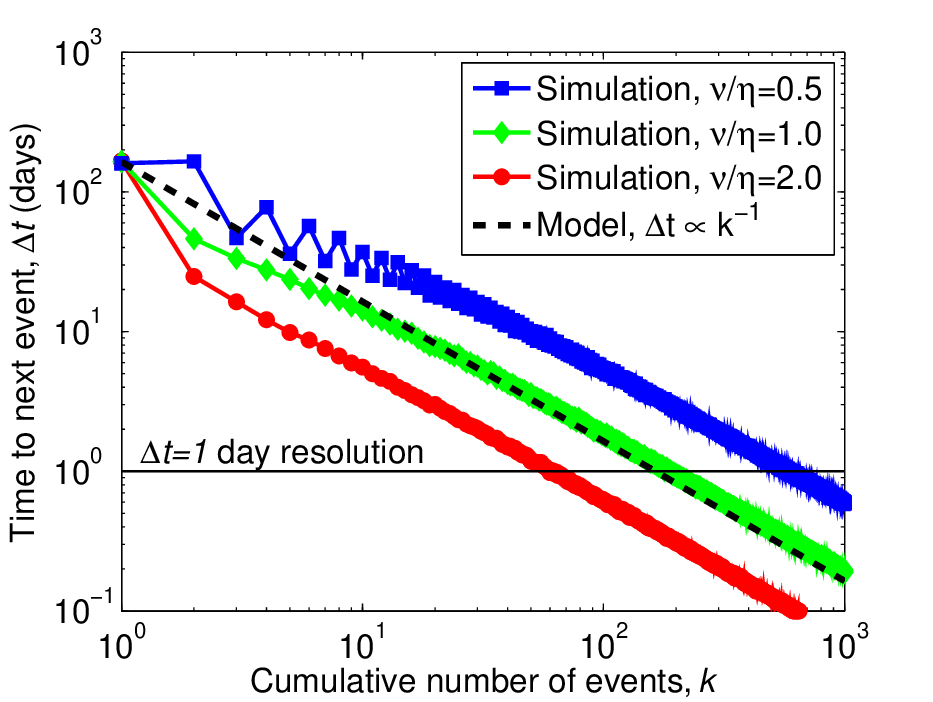}
\includegraphics[scale=0.435]{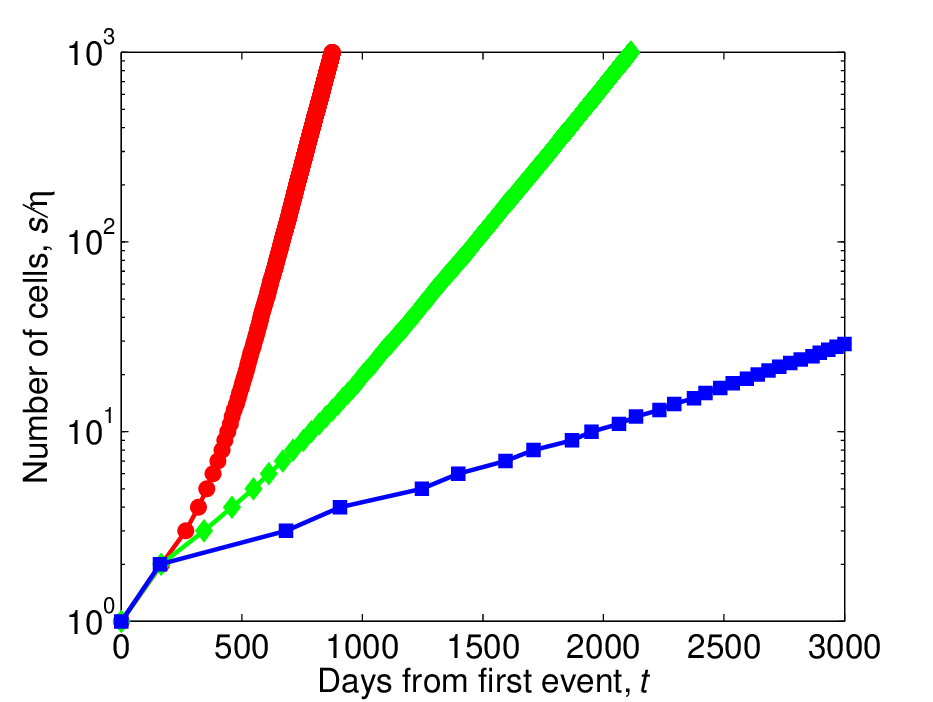}
\end{center}
\caption{\textbf{Simulated development of a terrorist organization.} (A) Median event delay~$\deltat$ vs.\ cumulative number of events~$k$, for 10,000 simulated terrorist organizations and three choices of the number of cells $\nu/\eta$ added per event. Dashed line shows the function $\deltat\propto k^{-1}$, from Eq.~\eqref{eq:model}. (B) Median size (number of terrorist ``cells'' $s/\nu$) vs.\ calendar time from the first event, showing exponential growth with rate set by $\nu/\eta$.}
\label{fig:simulation}
\end{figure}

This simple model intentionally omits many factors, such as
organizational structure, political motivation, geography, etc.,
that are likely to impact the behavior of any particular
organization. We also intentionally omit any potential response by
state-level actors and their consequences on the organization's
evolution. This last decision is made in order to focus on the
development of the organization, i.e., its early lifetime, where
labor constraints are likely most profound, although such
processes could naturally be added. Omitting these factors keep
the model simple and allows us to make quantitative predictions of
the generic relationship between organization size and the
frequency and severity of its attacks via direct numerical
simulation. To mimic the natural variation between particular
events, for each new event being planned by a cell, we draw a
delay $\tau$ from a fixed distribution. (In general, our results hold so long as the distribution of $\tau$ is well-behaved and stationary with respect to $k$.) Specification details and
computer code for the simulation are given in the Supporting Information.

Each simulated terrorist organization generates a unique sequence of events representing the collective behavior of its cells over time, and we extract the generic behavior by computing quantiles over variables of interest for many such simulated organizations. Here, we are interested in how the delay between subsequent attacks~$\deltat$ varies with cumulative number of events~$k$ (H3), and how the size of the organization, measured by the number of cells $s/\eta$ varies with calendar time~$t$ from the first event (H2). H4 predicts that event severity correlates with organization size and thus no additional information is gained by explicitly simulating event severities.

Figure~\ref{fig:simulation} shows the results for 10,000 simulated organizations, for three choices of the ratio $\nu/\eta$, which represents the growth rate of the organization's militant wing. When $\nu/\eta<1$ regime, organizational growth is slow because multiple events are required to establish a new terrorist cell; but, when $\nu/\eta>1$, organizational growth is fast because each event produces at least one new cell.

The generic behavior of our model is clear: (i) organizational size grows exponentially with time, at rate $\nu/\eta$, and (ii) the feedback between size and production rate induces a strong correlation between experience, size and the frequency of events. Finally, the model produces a universal functional relationship between delay~$\deltat$ and cumulative production~$k$ of the form~$\deltat \propto k^{-1}$, and this relationship is independent of the growth rate~$\nu/\eta$.

This latter point is worth reiterating: so long as each new event leads to some marginal increase in the overall production rate~(H2), a positive feedback loop between size and event production will exist. This feedback will be linear $\deltat\propto k^{-1}$ if the growth rate~$\nu/\eta$ does not vary with experience~$k$. If the militant wing is a decreasing fraction of the overall organization ($\nu/\eta$ decreases over time), the feedback will be sub-linear and $k^{-\beta}$ with $\beta<1$, while if it increases with time, the feedback will be super-linear and~$\beta>1$. These properties imply that if a growing organization does provoke responses from state-level actors, these responses will not break the feedback loop unless they succeed in both limiting the growth and reducing the size of the organization, a point to which we will return later.

These quantitative predictions can be tested with empirical data by examining~$\deltat$ as a function of~$k$ across many organizations. If $\deltat\propto k^{-1}$ holds in the data, we have strong evidence for precisely the size-mediated feedback loop described here.

\subsection*{Empirical data}
Organizational size data were drawn from the Big Allied And Dangerous (BAAD) data set~\cite{asal:rethemeyer:2008}, which offers the currently best available size estimates for terrorist organizations worldwide. Other sources of size data lack the breadth or temporal resolution for accurate analysis. For instance, the START program and the MIPT database previously held a small number of estimates of uncertain accuracy, generated by Detica, Inc., a British defense contractor, and \cite{jones:libicki:2008} compiled a database of information on 649 terrorist groups that included only estimates of the maximum size over a group's entire lifetime. The BAAD data were generated by a survey of domain experts at the Monterey Institute of International Studies (MIIS) who estimated the rough order of magnitude (1--100, 100--1000, 1000--10,000 and $>$10,000 personnel) of the maximum size achieved by each of 381 groups, between 1998 and 2005, identified in the~\cite{mipt:2008} event database. Of these, 161 organizations conducted at least one deadly attack, and 80 conducted at least two in that period.

To ensure good compatibility with this organization list, event data were drawn from the MIPT Terrorism Knowledge Base~\cite{mipt:2008}, which contained 35,668 terrorism events, of which 13,274 resulted in at least one fatality, as of 29 January 2008. 
(Other sources of event data include the Global Terrorism Database~\cite{gtd:2010}, the Worldwide Incident Tracking System~\cite{nctc:2009} and the ITERATE data~\cite{mickolus:etal:2004}. We note that neither these nor the MIPT database provide complete and consistent worldwide coverage.)
For the period 1968--1997, the MIPT database includes mainly international events involving actors from at least two countries, while for 1998--2008 it includes both domestic and international events from much of the world.
(The MIPT data were originally drawn from the RAND Terrorism Chronology 1968--1997, the RAND-MIPT Terrorism Incident database (1998--Present), the Terrorism Indictment database (University of Arkansas \& University of Oklahoma), and DFI International's research on terrorist organizations. In 2008, however, the U.S.\ Department of Homeland Security discontinued its funding for the maintenance of the database in favor of the University of Maryland's START center's Global Terrorism Database~\cite{gtd:2010}.)
Each event is defined as an attack on a single target in a single location (city) on a single day. For example, the Al Qaeda attacks in the United States on 11 September 2001 appear as three events in the database, one for each of the New York City, Washington D.C. and Shanksville, Pennsylvania locations. Each record includes the date, target, city (if applicable), country, type of weapon used, terrorist group(s) responsible (if known), number of deaths (if known), number of injuries (if known), a brief description of the attack and the source of the information.

The organizations identified in the MIPT database are a superset of those contained in the BAAD data set, and we will use these additional data analyses that do not require size estimates. For each organization, we extracted the full sequence of its attributed or claimed events. This yields 10,335 events worldwide from 1968--2008 associated with 910 identifiable organizations. For each of the 1,204 events worldwide with unknown severity, we assign a severity of $x=0$ to preserve timing information. Further, because of the day-level temporal resolution of events in the database, multiple events on the same day by the same group have ambiguous ``delay'' (inverse frequency). We eliminate this ambiguity by aggregating such events into a single ``event day'' with severity equal to the sum of the component severities. This slightly reduces the number of events, mainly for the most active organizations late in their life history. As a consequence, the minimum resolvable delay in the database for two events by the same organization is $\deltat=1$ day.

\subsection*{Regression models}

Before analyzing the evolution of attacks by individual organizations we conduct static or cross-sectional regression analysis at the level of individual organizations. We examine the relationship between group size and attack patterns, in particular the delay between attacks, the experience of a group in terms of number of events, and the severity of attacks.

To recap, we expect larger groups to generate a larger number of attacks, have shorter delays between attacks~(H1), and generate more severe attacks even accounting for other attack patterns~(H4). We can evaluate H1 by comparing maximum group size~$s$ from BAAD and the minimum delay between attacks~$\deltat$ in MIPT. We can assess H4 by comparing size and the maximum severity $x$ of attacks. Finally, H2 implies that larger groups should have higher maximum experience $k$ or cumulative number of events. (H3, postulating a declining delay with subsequent attack, cannot be evaluated with static data; we return to this point later.)

Although group size should predict attack patterns, individual
measures such as maximum severity will be at least in part a
function of the total number of attacks. That is, for any
distribution of severities, an increased production rate (sampling
intensity) will naturally inflate the maximum severity over a
fixed time period, even if the distribution is stationary. Thus,
in order to examine the partial relationship between size and the
related attack variables---or their independent predictive value
on size once we take into account the other attack pattern
characteristics---it is more convenient to consider to what extent
we can account for size as function of the attack measures.

We use an ordered logit regression model of size since the BAAD data give order-of-magnitude estimates of maximum size. As the BAAD data pertain to the time period 1998--2005, we restrict our attack pattern measures to attacks during this same time period. Since the distributions of minimum delay, maximum experience, and maximum severity are all highly skewed we take the natural logarithm, adding 1 to severity to prevent taking the log of 0 in the case of non-fatal events. We report the empirical estimates in Table~1.

\begin{table}[b]
\caption{Ordered logit regression of group size, by fatal attack patterns}
\vspace{2mm}
\label{table:OR:size} \centering
\begin{tabular}{|l|cc|} \hline
Variable & $\hat{\beta}$ & SE($\hat{\beta}$) \\ \hline
Delay: ln min$(\deltat)$ & -0.351 & 0.119 \\
Experience: ln max$(k)$ & 0.707 & 0.193  \\
Severity: ln max$(x)$ & 0.150 & 0.159  \\
$\hat{\alpha}_{0|1}$ & -0.163 & 0.840  \\
$\hat{\alpha}_{1|2}$ & 2.652 & 0.895   \\
$\hat{\alpha}_{2|3}$ & 5.039 & 1.056   \\ \hline
\multicolumn{3}{l}{N = 80, LR $\chi^2$ = 41.42, df = 3, 58.75\% correctly classified} \\
\end{tabular}
\end{table}

The results display a significant negative relationship between
fatal attack delay and group size, consistent with our claim that
larger groups will have shorter delays between attacks~(H1). We
also find a positive relationship between group size and
experience, consistent with our claim that larger groups generate
a higher number of attacks~(H2). Finally, the maximum severity of
the attacks is not significantly related to group size, once we
have controlled for delay and experience variables. This
contradicts the hypothesis that larger groups are systematically
more likely to generate severe attacks~(H4). Overall, the model
places 58.75\% of all the groups in the correct bins for group
size. Only 5\% of the observations are badly mis-classified, with
predictions off by more than one order of magnitude. By contrast,
a null model predicting all groups to have the modal size category
($100-1000$) only classified 43.75\% of the observations correctly.
(We considered a number of alternative specifications. Severity remains an insignificant predictor of group size when we consider combinations of delay and experience for both deadly and non-deadly attacks. Using a linear regression model rather than ordered logit does not change our substantive conclusions.)

\begin{table}[t]
\caption{Linear regression of experience, by attack delay and severity}
\vspace{2mm}
\label{table:glm:k} \centering
\begin{tabular}{|l|cc|cc|} \hline
& \multicolumn{2}{c|}{Fatal attacks ($F$)} & \multicolumn{2}{c|}{All attacks ($A$)} \\
Variable & $\hat{\beta}$ & SE($\hat{\beta}$) & $\hat{\beta}$ &
    SE($\hat{\beta}$)  \\ \hline
Delay$_F$: ln min$(\deltat)$ & -0.119 & 0.042 &  -0.110 & 0.040 \\
Delay$_A$: ln min$(\deltat)$ & -0.778 & 0.110 & -0.795 & 0.105 \\
Delay$_F \times$ Delay$_A$ &  0.074 & 0.017 & 0.073 & 0.016 \\
Severity: ln max$(x)$ &  0.190 & 0.059 &  0.150 & 0.056 \\
$\hat{\alpha}$ &  3.115 & 0.236 &  3.336 & 0.225 \\ \hline
& \multicolumn{2}{l|}{N = 167, R$^2$ = 0.545} & \multicolumn{2}{l|}{N = 167, R$^2$ =  0.565}  \\
& \multicolumn{2}{l|}{R$^2$ ($\neg$severity) = 0.515} & \multicolumn{2}{l|}{R$^2$ ($\neg$severity) = 0.546}  \\
& \multicolumn{2}{l|}{R$^2$ ($\neg$delay) = 0.222} &
    \multicolumn{2}{l|}{R$^2$ ($\neg$delay) = 0.182}  \\ \hline
\end{tabular}
\end{table}

Since the BAAD data cover only about half of the identifiable organizations in the MIPT database over a restricted time span (1998--2005), we conduct a supplementary analysis with the full MIPT dataset, where we consider how a group's total experience can be accounted for by differences in minimum delay and maximum attack severity.
(We limit the analysis to MIPT organization that generated at least two events (frequency) and one deadly event (severity); only 167 organizations satisfy these criteria.)
Table~2 report the results for a linear regression with logged
values for all the terms for fatal ($F$) and all attacks ($A$,
including non-fatal attacks) experience respectively. The results
clearly show that the minimum delay is a significant predictor of
group experience, and they mildly support the claim about
severity, as the positive coefficient for severity is
significantly different from 0. However, comparing the change in
the R${}^{2}$ for estimating the model with and without the
severity and delay terms respectively indicates that dropping the
severity variable leads to a relatively small decline, while the
impact of omitting the delay variables is substantial. Hence,
variation in delay between attacks accounts for much more of the
variation in experience than does severity.

These static analyses provide substantial preliminary evidence in support of H1 and H2 and little evidence to support H4. We now go beyond static analyses and test our predictions for all organizations in the MIPT database using a novel dynamical analysis tool called a ``development curve.''

\subsection*{Developmental dynamics}
A development curve is a statistical tool that measures the evolution of organization behavioral variables along a common quantitative timeline~\cite{clauset:gleditsch:2009}. It is similar in structure and use to the ``experience'', ``learning'' and ``progress curves'' sometimes used in management science~\cite{dutton:thomas:1984,thompson:2010} to quantify the relationship between per-item production cost (or time) and ``experience'' (cumulative item production). Because we study behavioral variables rather than the costs of production, and to explicitly avoid implying learning-based mechanisms, we choose a distinct term. The analysis of these developmental curves facilitates direct comparisons of the behaviors of different groups at similar points in their life histories, which is useful for testing our hypotheses.

We instrument a common timeline using organizational experience $k$, defined as the cumulative number of events produced by or associated with a particular organization, and we compare the delay $\deltat$ between the $k$th and $(k+1)$th events, or the severity $x$ of the $k$th attack, across all organizations in our sample. For each of the 910 organizations, we extract from the MIPT event data an ordered sequence of coordinates $\{(1,z_{1}),(2,z_{2}),\dots\}$, which represent the group's behavioral trajectory on the variable $z$ over its lifetime. The visualization of such trajectory is typically made using double-logarithmic axes, as illustrated in our simulation results in Figure~\ref{fig:simulation}. Although the curve construction itself ignores details such as the date of an organization's first attack, its location, ideology, etc., these variables can be used for subsequent analysis, e.g., comparing the trajectories across covariates.

Constructing a development curve for an individual organization (see Supporting Information) can facilitate the investigation of specific behavioral dynamics of individual groups over their lifetimes. However, the specific factors associated with particular organizations may obscure the generic tendency embodied by our hypothesis. To investigate these, we examine the average trajectory across many organizations by tabulating the conditional distribution~$\Pr(\deltat\,|\,k)$ of delays, for a specified level of experience~$k$. Thus, an organization that has carried out $k^{*}$ events contributes to each of the $k\leq k^{*}$ conditional distributions. This approach provides a strong test of the frequency-acceleration (H3) and attack-severity hypotheses (H4) predictions.

\subsubsection*{Frequency of attacks over time}
Figure~\ref{fig:frequency:curves}A shows the composite frequency curve for all organizations in our study. To reduce the overprinting effects of showing the trajectories for so many organizations, we bin the values of~$k$ on a logarithmic scale and plot the mean and 1st and 3rd quartiles of the data within each bin. Remarkably, the observed empirical pattern agrees very closely with our simulation model's predictions (Figure~\ref{fig:simulation}).

The progressive decrease of the delay distributions indicates a generic tendency toward faster production with increased experience for all types of organizations, in strong agreement with the frequency-acceleration hypothesis (H3). But, the relationship between delay and experience is not deterministic: not every event occurs more quickly than the last but the statistical tendency toward shorter delays is clear.

A terrorist organization thus typically begins in the low-frequency domain (large $\deltat$) and moves in fits and starts toward the high-frequency domain (small $\deltat$). This trend is not subtle: the median delay after the $1$st event is $\deltat=124$~days, while by the $12$th event, it has dropped to $35$~days and by the $25$th, the next event typically comes only $21$~days later. This transition to fast production does take considerable calendar time: for groups that achieve $k=12$ events, the median total calendar time between the first and twelfth event is $4.4$~years. Similar results hold for the timing between deadly attacks.

None of the sampled organizations progressively slowed their attack rate over time, moving from high-frequency to low-frequency. A few unusual groups, such as Al-Qaeda in the Land of Two Rivers, begin and remain in the high-frequency domain. But, Al-Qaeda in the Land of Two Rivers is an interesting case because it is well-known to have operated under a different name prior to 2004~\cite{fishman:2008}; thus, their initial high-frequency behavior can be interpreted as support for the labor-constraint hypothesis (H1) because their initial larger size---a hold over from their previous identity---allowed them to ``begin'' life ($k=1$) at a relatively high initial production rate of attacks.

\subsubsection*{Statistical model for the frequency of attacks}
Quantifying the dynamical relationship between delays and experience allows us to go beyond our static analyses. To do this, we statistically model the conditional distribution $\Pr(\deltat\,|\,k)$ from which delays are drawn and how this distribution varies with experience.

For these data, a truncated log-normal distribution, with the following mathematical form
\begin{align}
\Pr(\deltat\,|\,k) & \propto {\rm exp}\!\left[\frac{-(\log \Delta t+\beta \log k-\mu)^{2}}{2\sigma^{2} }\right]  \enspace ,
\label{eq:model}
\end{align}
provides an excellent fit to the empirical delay data for all organizations.  Here,~$\sigma^{2}$ is the variance in delays at a given~$k$, $\mu$~is related to the characteristic delay between attacks and $\beta$ controls the rate at which that delay decreases with increased experience~$k$. That is, $\beta$ governs the strength of the feedback loop between organizational experience and the production of new events. To include the effect of the minimum timing resolution $\deltat\geq1$ present in the empirical data, we force $\Pr(\deltat\,|\,k)=0$ for $\deltat<1$~day.

\begin{figure}[t]
\begin{center}
\includegraphics[scale=0.435]{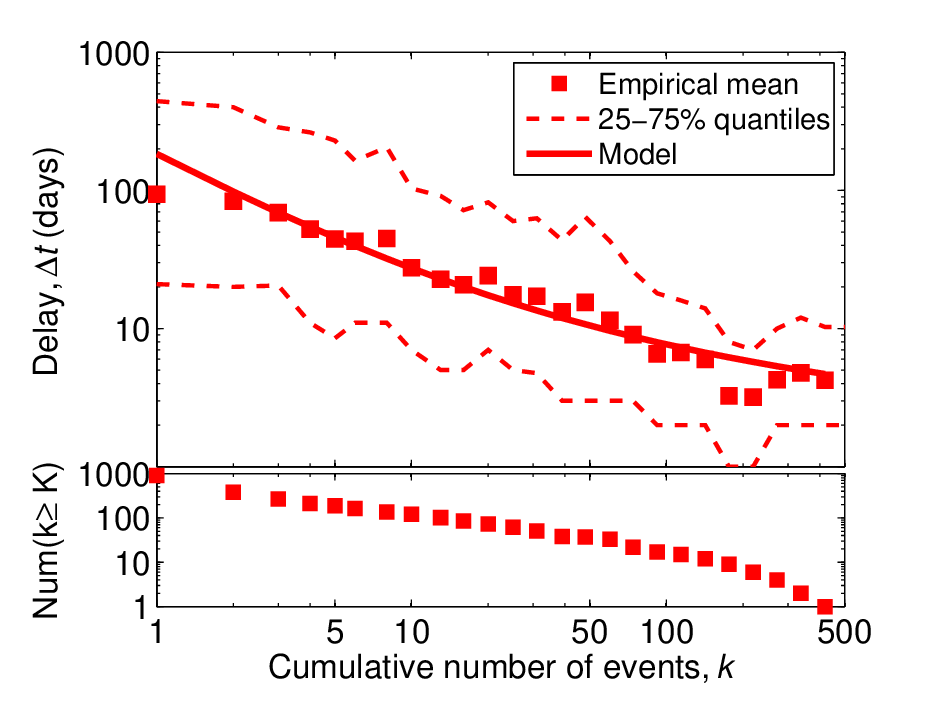}
\includegraphics[scale=0.435]{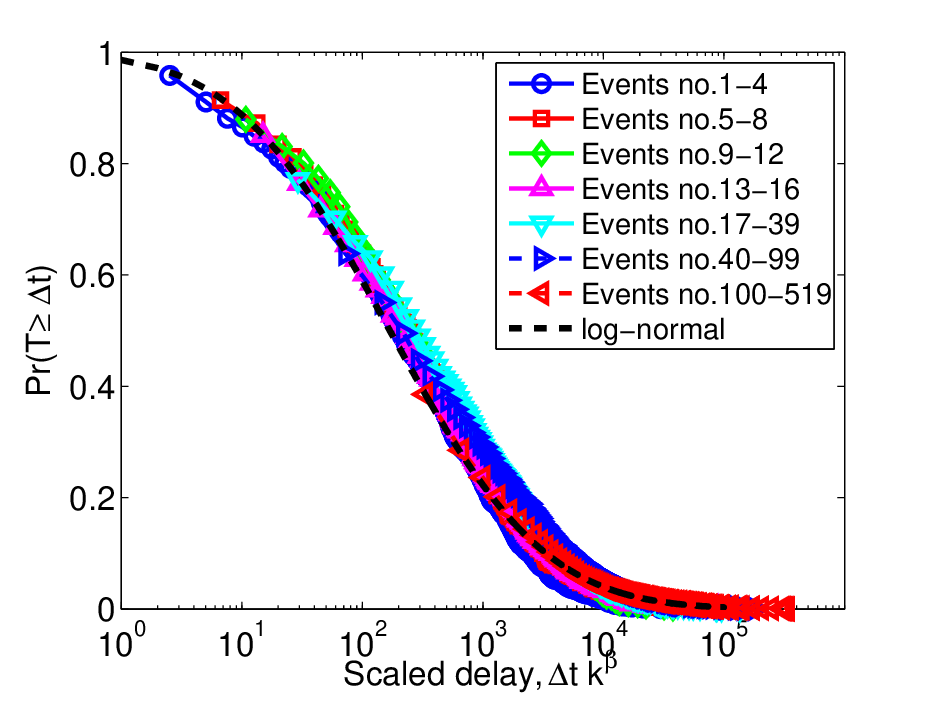}
\end{center}
\caption{\textbf{Timing of events.} (A) Mean delay~$\langle\log\deltat\rangle$ between attacks, with 1st and 3rd quartiles, vs.\ group experience~$k$. Solid line shows the expected mean delay, from the statistical model described in the text. Lower panel shows the number of organizations with at least $k$ evens. (B) A ``data collapse'' showing the alignment of the re-scaled conditional delay distributions~$\Pr(\deltat\cdot k^{\hat{\beta}}\,|\,k)$ with the estimated underlying log-normal distribution, as predicted by the model.}
\label{fig:frequency:curves}
\end{figure}

This mathematical structure implies that the typical delay between attacks generically decreases according to a power-law function with increasing experience
\begin{align}
\deltat\approx\rm{e}^{\mu}\,\!k^{-\beta} \enspace .
\label{eq:model:appx}
\end{align}
(Details of this derivation are given in the Supporting Information.)
Thus, if $\beta>0$, we will observe a transition toward increasingly fast event production, indicating support for H3. In contrast, if $\beta=0$, production rates do not vary with organizational experience, while if $\beta<0$, production rates will decrease (larger~$\deltat$) with increasing experience. In the $\beta>0$ regime predicted by H3, the acceleration effect is dampened as the mean delay asymptotes to the minimum timing resolution at~$\deltat=1$; this produces slight upward curvature for large values of $k$ (see Supporting Information).

The particular value of $\beta$ has a strong effect on the material dynamics of the feedback loop between increasing experience and increasing production. If $\beta=1$, then the feedback loop is linear, as in our simulation model, and increases in organizational experience lead to proportional increases in event production. Linearity implies that the marginal growth associated with an additional event is relatively constant over the organization's lifetime and a roughly constant fraction of new recruits are allocated to increase overall tempo of militant activities.

In contrast, $\beta\not=1$ implies a non-linear feedback process. Notably, non-linear feedback processes are not common models of social processes (but see the literature on arms-races, particularly~\cite{richardson:1960} and~\cite{wallace:wilson:1978}). Traditional models often focus on proportional effects in which increases in one variable cause proportional changes in other variables. In non-linear feedback processes, small increases in one variable can produce dramatic and continuing swings in other variables, leading to highly unpredictable dynamics~\cite{strogatz:2001}.

When $\beta>1$, the feedback is super-linear, and one or both of these factors must increase with~$k$. That is, either per-event growth in militant activities increases over time or an increasing fraction of growth is allocated to militant activities. When $\beta<1$, the feedback is sub-linear and the marginal recruitment benefits of new events decrease over time or they are constant but recruits are increasingly allocated toward non-militant activities.

Fitting this model directly to the empirical data on all events, we find that the maximum likelihood estimate is~$\hat{\beta}=1.0\pm0.1$ (std.\ err.), indicating linear feedback. (This approach to estimating the parameter gives weight to the events early in organization's lifetime that is proportional to the number of such events in our data set; in contrast, a simple regression approach on the mean delays would bias the estimate by giving significant weight to the rare but long-lived groups.) Using a Monte Carlo simulation against a null model with fixed $\beta=0$ (no acceleration over time) and with $\mu$,$\sigma$ estimated using maximum likelihood given the fixed $\beta$ value, we find that the value of $\hat{\beta}$ is highly statistically significant ($p<0.001$). (Fitting to deadly attacks alone yields a highly statistically significant~$\hat{\beta}=1.1\pm0.2$, slightly in the super-linear regime, but this value is statistically indistinguishable from $\beta=1$.)

A linear feedback implies that the marginal growth from event-driven recruitment does not vary much with organizational size or experience. Furthermore, it implies that organizational learning in terrorist groups~\cite{jackson:etal:2005,johnson:etal:2011}, in which the production rate increases due to improved efficiency of a fixed number of individuals, plays a lesser role in explaining the overall acceleration of event production than do the effects of increasing organizational size, because learning would mimic the effect of super-linear feedback by allowing a constant number of militants to behave identically to an increasing number.

A strong test of the statistical model's plausibility is its prediction that each of the $k$ conditional delay distributions $\Pr(\deltat\,|\,k)$ is a scaled version of the underlying log-normal distribution LN($\mu,\sigma^{2})$. To test this prediction, we re-scale the empirical distributions by the predicted factor, i.e., we multiply each delay variable $\deltat_{i}$ by $k_{i}^{\hat{\beta}}$, and then plot them against the estimated underlying log-normal distribution. A close alignment of these re-scaled conditional distributions, also called a ``data collapse''~\cite{bhattacharjee:seno:2001}, is strong evidence for the hypothesized data model over a wide range of alternatives. Furthermore, for an alternative model to produce such a data collapse requires that it follows the log-normal form closely enough to be effectively equivalent.  Figure~\ref{fig:frequency:curves}B shows the results of this test, illustrating an excellent data collapse, with each of the re-scaled log-normal conditional distributions closely aligning with the underlying log-normal form.

\begin{table}[t]
\caption{Frequency curve parameters for organizations with similar political motivations. Note: statistical significance estimated via Monte Carlo simulation of a two-tail test against a null model with $\beta=0$ (no frequency acceleration), using the sum-of-squared errors (SSE). Values in parentheses indicate bootstrap standard uncertainty in the last digit.}
\vspace{2mm}
\label{table:frequency:models}
\centering
\begin{tabular}{r|cc|ccc|c}
political motivation & groups & events & $\mu$ & $\sigma$ & $\beta$ & significance\\
\hline
nationalist-separatist & 55 & 2959 & $5.1(5)$ & $2.2(1)$ & $0.9(2)$ & $p<0.001$ \\
reactionary & 5 & 143 & $3.2(6)$ & $1.8(2)$ & $0.1(3)$ & $p<0.001$\\
religious & 17 & 999 & $5(1)$ & $2.4(5)$ & $1.7(5)$ & $p<0.001$\\
revolutionary & 53 & 2527 & $5.7(4)$ & $2.3(2)$ & $1.1(2)$ & $p<0.001$\\
\hline
all secular & 883 & 6232 & $5.2(2)$ & $2.25(9)$ & $0.9(1)$ & $p<0.001$ \\
\hline
all groups & 910 & 7231 & $5.1(2)$ & $2.32(9)$ & $1.0(1)$ & $p<0.001$
\end{tabular}
\end{table}

These results also hold when we consider the development curves for groups with a common political ideology (see Supporting Information). \cite{miller:2007} divides the political motivations for terrorism into four conventional categories: nationalist-separatist, reactionary, religious and revolutionary. We coded according to Miller's criteria the 131 most prolific groups in our sample (all with $k\geq10$ deadly events), which accounts for 85\% of events, and fitted Eq.~\eqref{eq:model} to the data within each ideological category. Organizations with multiple political motivations were placed in multiple categories, which would only lessen any differences between estimated parameters for different categories. Within each of these categories, we observe the same acceleration pattern, with the strongest acceleration (largest~$\beta$) appearing for religious groups (Table~\ref{table:frequency:models}).

\subsubsection*{Severity of attacks over time}
\begin{figure}[t]
\begin{center}
\includegraphics[scale=0.435]{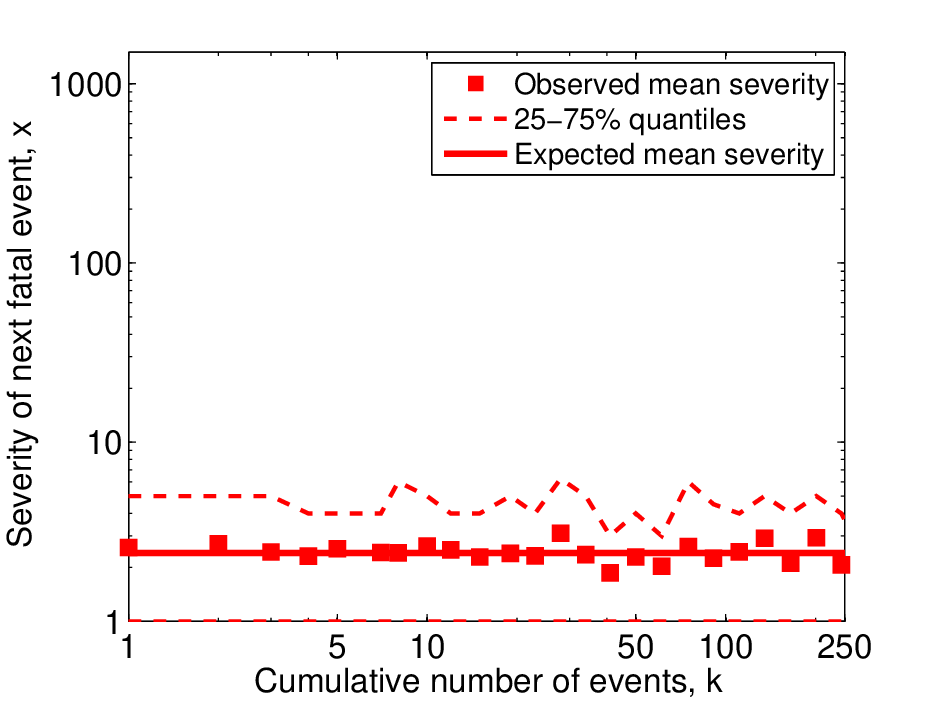}
\includegraphics[scale=0.435]{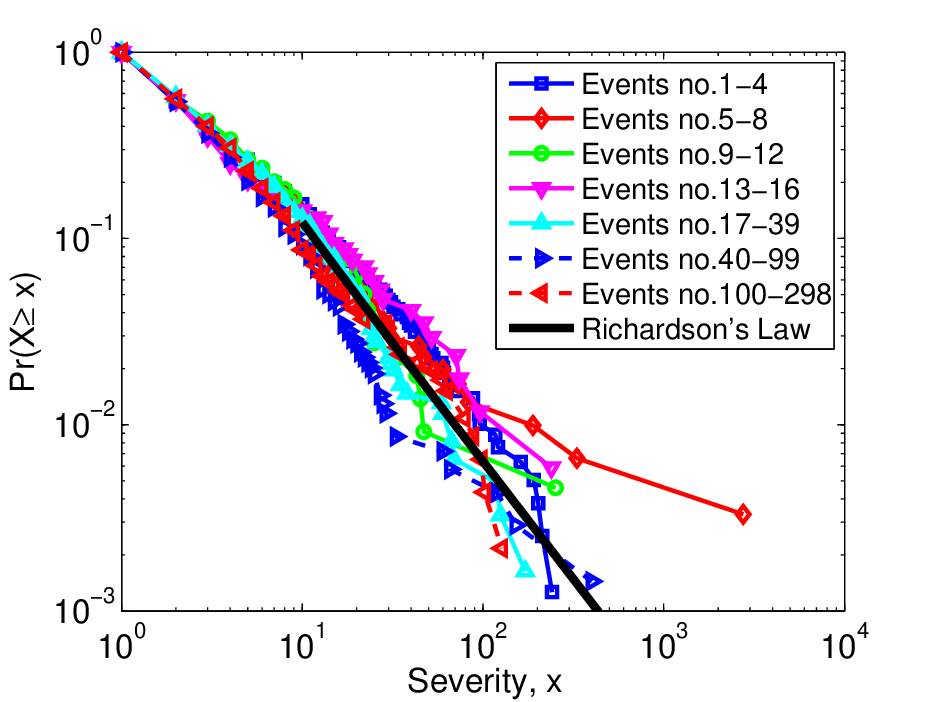}
\end{center}
\caption{\textbf{Severity of events.} (A) Mean severity~$\langle\log x\rangle$ of deadly attacks, with 1st and 3rd quartile, vs.\ group experience~$k$. Solid line (with slope zero) shows the expected delay, from a simple regression model. (B) Conditional severity distributions~$\Pr(x\,|\,k)$, showing a data collapse onto a heavy-tailed distribution, with the maximum likelihood power-law model for all severities (Richardson's Law).}
\label{fig:severity:curves}
\end{figure}

In contrast to the delay development curve, we find no statistically significant relationship between the severity of attacks and increased experience (Pearson's $r=-0.024$, t-test,~$p=0.17$), indicating no support for the severity-increase hypothesis (H4). Across all organizations in our sample, the average severity of the first deadly event is~$\langle x\rangle=6.7\pm0.9$, which is only slightly larger than the average severity of deadly events by highly experienced groups (those with $k>100$)~$\langle x\rangle=5.1\pm0.6$. Figure~\ref{fig:severity:curves}A shows the composite severity curve for all organizations in our study.

As with the frequency curves, we find that the conditional
severity distributions~$\Pr(x\,|\,k)$  roughly collapse onto a
single, underlying form (Figure~\ref{fig:severity:curves}B), which
is similar to the power law observed for all deadly terrorist
attacks worldwide from
1968--2008~\cite{clauset:etal:2007,clauset:etal:2009}. That
is, Richardson's Law for terrorism appears to hold for both
inexperienced and highly experienced groups. Combined with our
static analysis of organizational size, this pattern implies a
highly counter-intuitive fact: the severity of attacks by larger,
more experienced organizations, is not significantly greater than
the severity of attacks by small, inexperienced organizations.
That is, the common assumption that only experienced groups are
capable of such mass destruction~\cite{jordan:2009} is incorrect:
inexperienced organizations are just as likely to produce
extremely severe events as highly experienced organizations.

However, although more experienced organizations are not systematically more lethal at the individual-event level, the observed frequency-acceleration pattern implies that more experienced groups are significantly more lethal overall. This pattern was observed by~\cite{asal:rethemeyer:2008} in their analysis of the BAAD organizations. Our results thus clarify their results, showing that the observed correlation between greater lethality (total deaths attributed to an organization) and greater organizational size appears because larger, more experienced organizations produce events more quickly than smaller, less experienced organizations. It is the cumulative effect of the many small events that generates an increased lethality, not a systematic increase in the lethality of individual events.

Repeating this analyses on our ideology-coded set of organizations, we find no systematic dependence of severity of attacks on organizational experience within any of the ideological categories (see Supporting Information). That is, none of the model coefficients are significant, and the average severity of events within each category vary only a little. In short, we find that political ideology has no systematic impact on the severity of events or the trajectory that event severities take over the lifespan of an organization.

\section*{Discussion}
Although details and circumstances vary widely across terrorist organizations, the generic nature of our results suggests general conclusions. In particular, we find strong evidence for a positive feedback loop among organizational size (number of personnel), experience (cumulative number of events) and the frequency at which that organization launches new events. Small and inexperienced organizations tend to produce events slowly, while larger and more experienced organizations tend to produce events sometimes hundreds of times more frequently.

Within this feedback loop, new attacks lead to organizational growth and the corresponding increase in size leads to faster production of new events because a larger size means more terrorist cells are operating in parallel, not because events themselves are planned more quickly. The result of this feedback loop is a generic ``developmental'' trajectory: as an organization ages, it tends to produce violent events more and more quickly.

The typical form of this relationship can be mathematically modeled by a power-law function, in which the delay $\deltat$ between consecutive events decreases roughly like $\deltat\propto k^{-\beta}$ where $k$ counts the cumulative number of events and $\beta$ describes the strength and direction of the feedback loop. The implication of the power-law pattern is that large organizations are very much like ``scaled up'' versions of small organizations, and in particular that size and experience are coupled in a positive feedback loop.

Across all organizations in our sample, we estimate $\beta=1.0\pm0.1$, indicating a linear feedback loop, which implies that an organization's overall size is strongly correlated with the size of its militant wing. This pattern is strongest for small or inexperienced organizations, e.g., those with $k\leq 10$ events, which covers 87\% of the 910 organizations in our sample. In contrast, highly experienced organizations seem to saturate their event production rates at the daily or weekly level, which may be indicative of a tendency of large organizations to engage in multiple types of activities, e.g., the provision of social services, criminal activities, etc., continuing to grow their militant wings.

The mathematical precision of this relationship is striking, as is the ability of our computer simulation to reproduce it. Except for Richardson's Law for the frequency and severity of wars, few statistical relationships in the study of political violence exhibit such regularity.

The power-law relation between organizational experience and production rate is both conceptually and mathematically similar to the relationship between cost and cumulative production observed in manufacturing~\cite{dutton:thomas:1984} or organizational learning~\cite{argote:1993,thompson:2010}, where decreases in per-item production costs or time can be described by a power law in the cumulative number of items produced. That a similar patterns appears in the production of terrorist events is surprising, and it may not be superficial to describe terrorist organizations as a special type of manufacturing firm whose principal product is political violence and whose overall production of violence is fundamentally constrained by its size.

The implication is that terrorism is inherently non-amenable to mass production, i.e., it is not a scalable enterprise, perhaps because each event must be humanly conceived and planned around a particular target, tactic or environment, and there is a limit to how much this process can be automated. One implication of this conclusion for cyber-terrorism is that even there, despite the great potential for automating attacks, these too will likely not be scalable without advances in general artificial intelligence.

In the language of economics, we say that terrorism capital and labor are not freely substitutable with respect to producing new events. If the day-to-day work of event production does not require specialized skills, then the growth potential of an organization be extremely large because it may draw on the largest possible pool of potential recruits. This point suggests that conflict-level event production rates should ultimately be responsive to policy and counter-terrorism efforts that target the size and mobility of the pool of potential recruits. That is, successful ``hearts and minds'' strategies~\cite{howard:2002} are likely to lead directly to lower incident rates by both restricting the growth and reducing the size of terrorist organizations. They may not, however, eliminate the possibility of spectacular attacks as these do not depend on organizational size.

Recently, following our original work on progress curves in terrorism~\cite{clauset:gleditsch:2009}, Johnson et al.~\cite{johnson:etal:2011} analyzed the timing of events in the Iraq and Afghanistan conflicts, finding similar power-law like acceleration curves in the delay between events. They argue that this pattern is caused by a kind of ``red queen'' effect---a concept borrowed from arms races in evolutionary biology~\cite{vanvalen:1973}---in which two sides of the conflict race through some abstract space, and the timing between events is given by how far ``ahead'' the insurgent side is in the race. In practice, however, this explanation is difficult to validate because the connection is not specified as to how real-world events and structures drive the dynamics of the abstract race. In contrast, our explanation of the phenomena is both tangible, general and testable: we argue that the size of the insurgency or the terrorist group sets the tempo of the conflict. The more people there are fighting, the more frequently we will observe events. This explanation makes direct and testable predictions about the relationship of organizational size and frequency of events, which we show are upheld by empirical data on organizational sizes. (As a technical note, in the language of physics, the ``size'' of an organization or insurgency is an extensive variable of the conflict system, much like area and number of particles are for physical systems~\cite{pathria:1996}; this fact makes additional testable predictions of our theory.) The implication for the Iraq and Afghanistan conflicts is that the number of insurgents active in the various provinces is the primary determinant of the frequency of events observed there.

Although the acceleration is remarkably strong, the vast majority of organizations do not achieve high levels of experience (only 23\% of groups are associated with $k>10$ events) or fast production rates. The progressive loss of organizations could be due to high rates of organizational death, e.g., from counter-terrorism activities or internal conflicts~\cite{cronin:2006,jones:libicki:2008}, shifts away from violence, or a right-censoring effect on young and still active organizations. Significantly, the particular mode of organizational demise seems not to have a strong impact on the production time of events, suggesting that the transition from development (growth) to death may happen very quickly, so that the experience curve does not bend upward but rather simply halts. Further exploration of the death of organizations~\cite{cronin:2006,jones:libicki:2008}, and how it impacts the production of violence, is an interesting avenue for future work.

Regardless of the reason, we do not expect the feedback loop to continue as $k\to\infty$. If an organization succeeds in becoming large enough to produce new events each day, it may function more like a stable or mature social institution, with fundamentally different constraints and incentives on the production of violence. Large size and stability may also pose special risks, e.g., leading to larger or longer conflicts. On the other hand, non-violent activities, e.g., engagement with political processes, may also become more attractive with increased size. Exploring these possibilities is an interesting avenue for future work. 

Unlike the production of events, we find no evidence of any relationship with the severity of attacks (H4). Rather, Richardson's Law---a power-law distribution in the frequency of severe events---characterizes the severity of events at all levels of organizational size or experience, and independent of the organization's political ideology.

This fact clarifies ongoing efforts to identify the underlying social, political or physical mechanism that generates Richardson's Law in terrorism. Several existing explanations assume or predict a severity-size relationship, e.g., the aggregation-disintegration model of Johnson et al.~\cite{bohorquez:etal:2009} and~\cite{clauset:wiegel:2010}, but these seem increasingly unlikely given our results here, because they assume the maximum severity of an event is proportional to the organization's size $N$; thus, if $N$ is small, the severity of events $x$ will also be small. That is, in their existing form, these models predict a severity-size relationship that does not appear in the data. Of course, these models may be adapted to produce the observed size-independence pattern, but doing so requires additional assumptions and additional validation that may not be warranted.

In contrast, two plausible explanations are not ruled out: (i) the explanation proposed in~\cite{clauset:etal:2007}, which posits a coevolutionary competition between states and terrorists in which event planning time and severity are strongly related, and (ii) the explanation proposed in~\cite{clauset:etal:2010:b}, in which population densities are broad-scaled and terrorists preferentially target high-density locations. Both of these explanations do not assume any relation between the severity of an attack and the size of an organization.

Together, our results suggest that the total lethality of larger and more mature groups observed by Asal and Rethemeyer~\cite{asal:rethemeyer:2008} is probably best explained as a natural consequence of their much more frequent activities, rather than as a systematic increase in the deadliness of individual events. Policies that limit the growth of an organization's militant wing should lower the long-term probability of a severe event by that organization. Such growth-limiting policies could be described as ``starving the beast'' of the labor necessary to produce rare but highly severe events.

The most productive targets of such policies will be large, established organizations with long histories of producing terrorist attacks. By virtue of their size, these organizations are likely to be well-known players in their particular conflicts and thus easy targets for specific policies. Because small organizations are equally likely to produce severe events, policies aimed specifically at large, well-known organizations may not limit the overall risk of severe events from all sources. For small and potentially unknown organizations, the most effective policies may be those aimed at preventing their formation in the first place, i.e., policies that curtail the acquisition of the means for and resort to violence. Lacking this, once such a terrorist cell carries out its first attack and begins its developmental trajectory, the best action by a government may be an ``overwhelming response'' to encourage through various means the dissolution of the nascent organization and the truncation of its growth trajectory. This policy is not without risk to the state, however, as certain countermeasures may serve the terrorist's goals~\cite{mueller:2006,ganor:2008}.

In closing, we point out that the acceleration in the frequency of terrorist events is independent of many commonly studied factors associated with terrorism, including geographic location, time period, international vs.\ domestic targets, ideological motivations (religious, national-separatist, reactionary, etc.), and political context. Our results thus demonstrate that some aspects of terrorism are not nearly as contingent or unpredictable as is often assumed and the actions of terrorists may be constrained by processes unrelated to strategic tradeoffs among costs, benefits and preferences. Identifying and understanding these processes offers a complementary approach to the traditional rational-actor framework, and a new way to understand what regularities exist, why they exist, what they imply for long-term social and political stability, e.g., large-scale violent conflicts like civil and interstate wars.

\section*{Acknowledgments}
The authors thank Lars-Erik Cederman, Konstantinos Drakos, Brian Karrer, James McNerney, John Miller, Mark Newman, Andrea Ruggeri, Didier Sornette, Cosma Shalizi, Brian Tivnan, Valerie Wilson and Maxwell Young for helpful conversations.
This work was supported in part by the Santa Fe Institute, the Economic and Social Research Council (RES-062-23-0259), and the Research Council of Norway (180441/V10).


\newpage

\section*{Supporting Information}

\renewcommand{\thefigure}{S\arabic{figure}}
\setcounter{figure}{0}
\renewcommand{\thetable}{S\arabic{table}}
\setcounter{table}{0}

\begin{itemize}
\item \textbf{Section~\ref{appendix:Z}}: \textit{Supplemental Analysis of Size, Frequency and Severity}: Additional analysis of the organizational size data, with respect to the frequency and severity of their events.
\item \textbf{Section~\ref{appendix:A}}: \textit{Development Curves for Four Prolific Organizations}: Individual frequency and severity development curves for the four most prolific organizations in the MIPT dataset.
\item \textbf{Section~\ref{appendix:B}}: \textit{Terrorist Organization Computer Simulation}: Specification and simulation code for the computer simulation described in the main text.

\item \textbf{Section~\ref{appendix:C}}: \textit{Statistical Model for the Frequency of Attacks }: Mathematical details of the statistical model for the generic pattern in event frequencies versus organizational experience.

\item \textbf{Section~\ref{appendix:D}}: \textit{Domestic vs.\ Transnational Events }:
Robustness check of the frequency acceleration pattern by considering organizations whose first event was prior to 1998 (mainly international terrorist organizations) versus after (mainly domestic terrorist organizations).

\item \textbf{Section~\ref{appendix:E}}: \textit{Political Ideology \& Frequency and Severity Curves }:
Variation in the developmental trajectories of organizations by political ideology, showing different frequency acceleration rates and no differences in event severity evolution.

\end{itemize}

\begin{appendix}

\section{Supplemental Analysis of Size, Frequency and Severity}
\label{appendix:Z}

The growth hypothesis predicts that a groupÕs maximum size will be inversely related to the minimum delay between its attacks over the 1998--2005 period. To complement the analysis in the main text, here we show the graphical plots and conduct additional analysis.

\begin{figure}[h!]
\begin{center}
\includegraphics[scale=0.445]{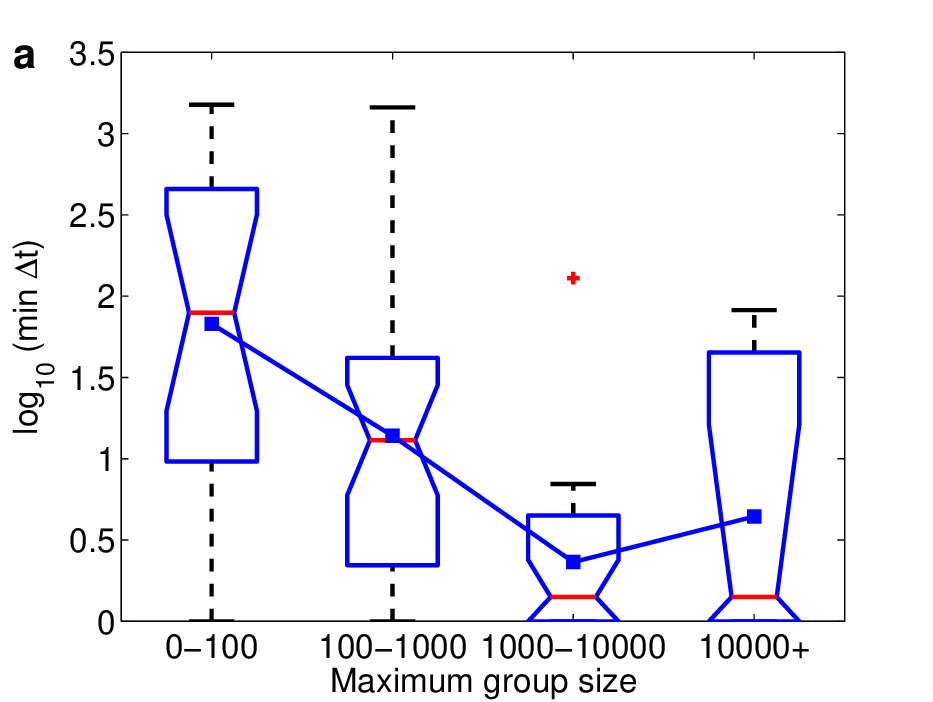}
\includegraphics[scale=0.445]{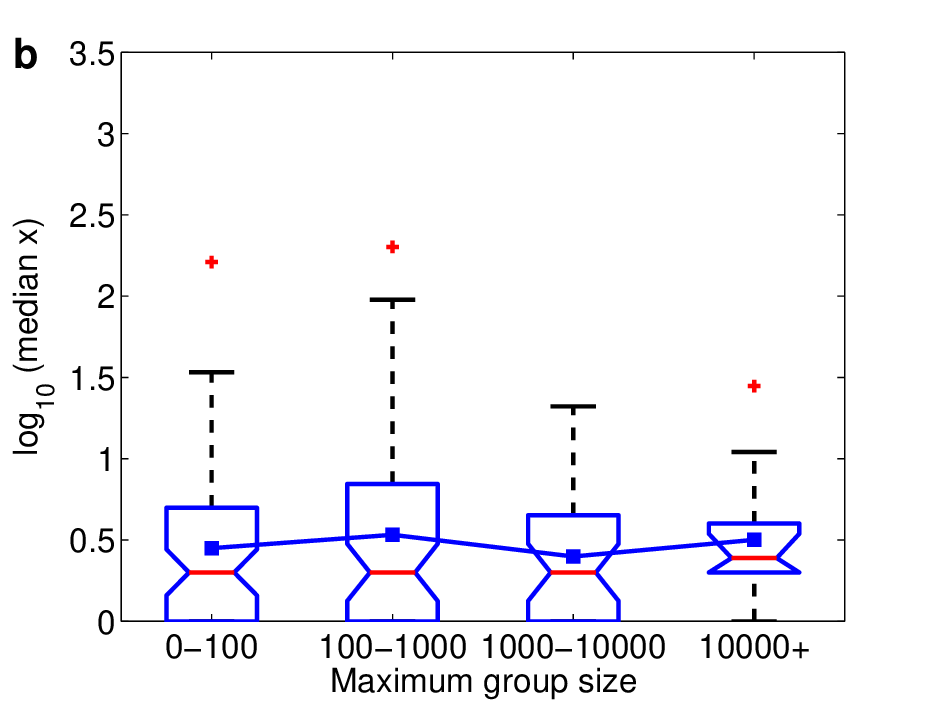}
\end{center}
\vspace{-4mm} \caption{Box-plots of the distributions of a groupÕs (a), minimum delay $\log(\min \Delta t)$ and (b), median attack severity $\log({\rm median} x)$ for attacks within 1998--2005, within each of four size categories. For convenience, we connect the means of each category, which are significantly different in the case of delays ($n$-way ANOVA, $F = 9.98$, $p < 0.000013$), but indistinguishable in the case of severities ($n$-way ANOVA, $F = 0.59$, $p = 0.62$). }
\label{fig:anovas}
\end{figure}

An analysis of variance indicates that the average minimum delays differ significantly between size categories ($n$-way ANOVA, $F = 9.98$, $p < 0.000013$). Further, we find that larger organizational size is a significant predictor of increased attack frequency ($r = -0.49$, $t$-test, $p < 10^{-5}$). Fig.~\ref{fig:anovas}a shows the distributions within the size categories. Although the distributions do overlap somewhat, the downward trend is clear.

In contrast, size, like experience, is not a significant predictor of median attack severity ($n$-way ANOVA $F = 0.59$, $p = 0.62$). Fig.~\ref{fig:anovas}b shows the distributions within the period. (We choose medians because they are robust to the large fluctuations caused by small samples drawn from heavy-tailed distributions.) Although there is some variability between size categories, the lack of a trend is clear.

\section{Development Curves for Four Prolific Organizations}
\label{appendix:A}

\begin{figure}[t]
\begin{center}
\includegraphics[scale=0.445]{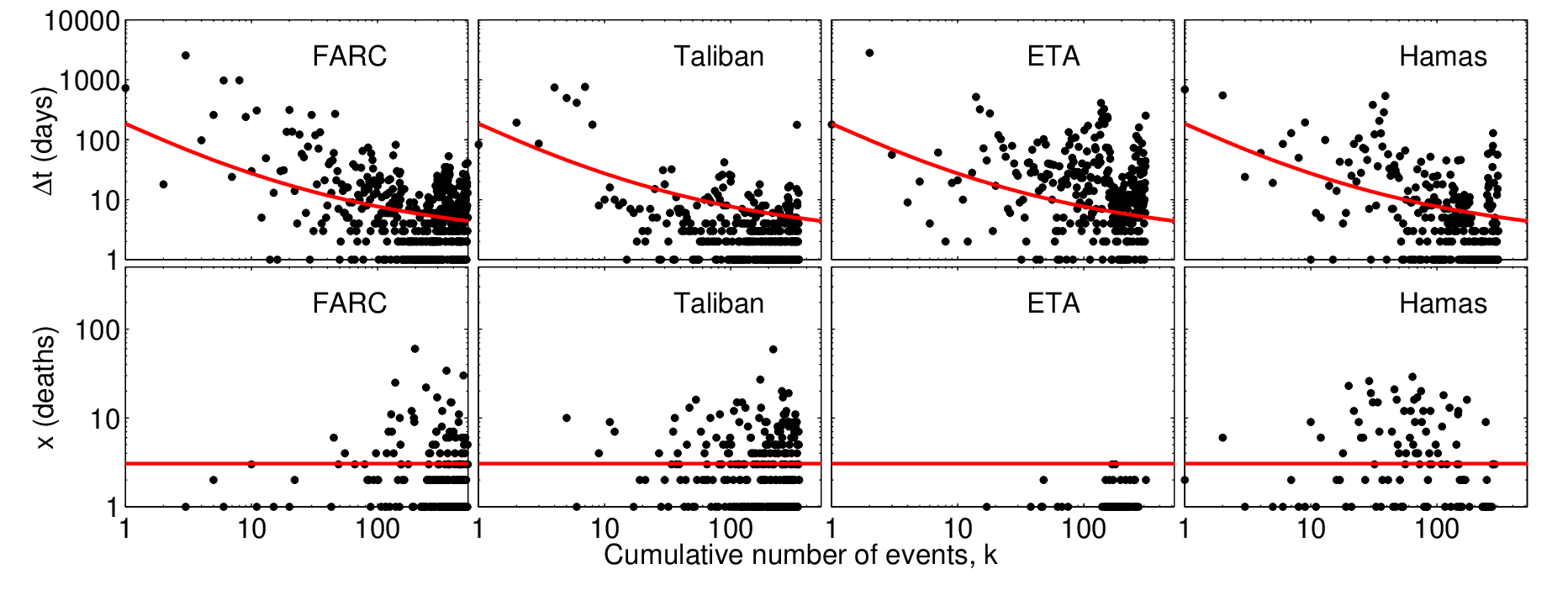}
\end{center}
\vspace{-4mm} \caption{Frequency (delay $\deltat$) and severity (deaths $x$) development curves for the Revolutionary Armed Forces of Colombia (FARC), Taliban, Basque Fatherland and Freedom (ETA), and Hamas, with generic trajectories estimated for all groups. Similar results hold for less experienced groups. }
\label{fig:individual:devcurves}
\end{figure}

As an example of development curve analysis, Figure~\ref{fig:individual:devcurves} shows the frequency and severity development curves for the four organizations with the greatest number of attributed event-days in our dataset, including both deadly and non-deadly events: the Revolutionary Armed Forces of Colombia (FARC; 520 events), the Taliban (349 events), Basque Fatherland and Freedom (ETA; 311 events), and Hamas (308 events). Non-deadly events ($x=0$) increment the counter $k$ for the severity curve but do not appear on the severity curve figures; hence, ETA, which carried out 261 (84\%) non-deadly events, has relatively few points in its severity curve.

For these organizations, the median delay between the $k=1$ and $k=2$ events is $\deltat=433$~days. In contrast, the median delay between the most recent pair of events by these groups is only $\deltat=4$~days, a 100-fold increase in frequency. In each case, the frequency curve begins in the upper-left corner of the figure, representing very long delays between subsequent events, and as $k$ increases, the curve moves consistently, albeit stochastically, toward the bottom-right corner, representing a convergence on very short delays between events.

This progression from slow to fast event production appears to
happen quickly: each of these groups achieves delays of
$\deltat\leq10$~days by their $k=12$th event. However, the median
calendar time required to achieve this high rate of production is
8.5 years; thus, although these first dozen events account for a
small fraction of the lifetime production of these organizations
(less than 4\% each), they account for a large fraction of the
organizations' overall lifetimes. Put more bluntly, these first
few events play a critical role in shaping the long-term
trajectory of an organization's production curve and they
illustrate a dramatic acceleration in the production of events as
the organizations mature. This important developmental effect is
obscured by high production rates later in life.

In contrast, the pattern for the severity development curve could not be more different: we observe no clear trend, either up or down, between event severity~$x$ and experience~$k$ for these organizations, and the median first and last severities are $x=0$ and $x=1$ deaths, respectively.
If anything, the only visual pattern we can discern is a possible increase in the variance of $x$ as $k$ increases. This preliminary analysis thus already indicates weak support for the severity-increase hypothesis (H4) but strong support for the frequency-acceleration hypothesis (H3). In combination with our static analysis above, this provides additional evidence supporting labor constraints and event-driven recruitment (H1 and H2).

\section{Terrorist Organization Computer Simulation}
\label{appendix:B}

The toy model described in the main text can be formalized and simulated explicitly. Below is computer code that implements the simulation in Matlab. In words, the simulation works as follows.

Let $\eta$ be a constant that denotes the number of individuals that make up a terrorist ``cell'' within the organization, and let $\nu$ be the number of individuals the organization as a whole gains via recruitment after each event. Thus, $\eta/\nu$ events are required to produce a single new cell; the particular values of $\eta$ and $\nu$ serve only to change the scale of the dynamics, not their fundamental character. Each cell is assigned a ``clock'' that measures the number of days remaining before that cell generates an event. We denote this delay $\tau$ and draw it from a log-normal distribution with parameters $\mu$ and $\sigma$, i.e., $\Pr(\tau)\sim \rm{LN}(\mu,\sigma)$. This is the only stochastic element of the simulation. When a cell generates an event, it then draws a new delay from the same distribution.

As described in the main text, each organization begins as a single cell, which has generated a single event at $t=0$. Thus, initially $s_{1}=\eta$. We then choose a delay $\tau$ for its next event. The simulation will generate a specified number of events, specified by the parameter $\verb+nok+$. For the $k$th event, the simulation then checks which cell has the smallest remaining delay and advances all cells' clocks by that much. It then generates the $k$th event, records its time as an ordered pair $(k,t_{k})$, and draws a new clock value for the generating cell. Additionally, it increments the organization's size by $\nu$ individuals, i.e., $s_{k}=s_{k-1}+\nu$, and adds $\lfloor s_{k}/\eta\rfloor$ new cells, each with a clock drawn from $\Pr(\tau)$.

A number of variations of this model generate equivalent results.
For instance, the distribution $\Pr(\tau)$ can generate very small
delays, e.g., less than 1 day, which may be considered
unrealistic. Imposing a minimum value on the $\Pr(\tau)$ does not
change the fundamental feedback between size and event production
and thus leaves the $k^{-1}$ trend unchanged. And, the ratio
$\eta/\nu$ only re-scales the underlying $k^{-1}$ behavior, as
seen in Figure~\ref{fig:simulation}. Finally, changing the
parameters of $\Pr(\tau)$ has no impact on the fundamental
behavior: the $\mu$ parameter sets the delay between the first and
second events, which appears as the expected $y$-intercept on the
resulting development curve, and varying $\sigma$ simply changes
the scatter around the underlying trend. In fact, the particular
functional form of $\Pr(\tau)$ we have chosen is not important,
and other choices lead to similar results; here, we choose the
log-normal distribution due to its similarity to the empirical
data (Fig.~\ref{fig:frequency:curves}).
\\

\baselineskip12pt
\begin{footnotesize}
\begin{verbatim}
% --- Terrorist organization simulation
% --- by Aaron Clauset

% --- set up simulation parameters
[mu  sigma] = deal(5.1,2.32); % parameters for Pr(tau) = LN(mu,sigma)
[eta nu]    = deal(5,5);      % size of cell,  marginal growth after an attack
nok         = 1000;           % number of events to generate

% --- set up simulation data structures
s   = zeros(nok+1,1); % organization size over time
c   = s;              % number of cells over time
[s(1) c(1)] = deal(eta,1);
fk  = zeros(nok+1,2);
fk(:,1) = (1:size(fk,1))';    % assign ids to events
gr      = zeros(nok+1,2);     % holds event clocks for each cell
gr(:,1) = (1:size(gr,1))';    % assign ids to cells

% --- initialize simulation: create the first cell
t   = 0;                      % global clock
k   = 1;                      % number of attacks to date (first attack at t=0)
tau = exp(sigma*randn(1)+mu); % choose delay from Pr(tau)
gr(1,:) = [1 tau];            % make first cell

% --- generate exactly nok events
while k<size(fk,1)

    % -- advance time to next attack
    [dt i] = min(gr(1:c(k),2)); % find cell with next attack
    t      = t + dt;            % advance all clocks by that much time
    gr(1:c(k),2) = gr(1:c(k),2) - dt;

    % -- generate the kth event
    k        = k + 1;           % increment attack number
    fk(k,2)  = t;               % record time of this event
    tau = exp(sigma*randn(1)+mu);
    gr(i,2) = tau;              % choose new delay for this cell

    % -- recruitment / growth
    s(k) = s(k-1) + nu;         % grow total personnel
    c(k) = floor(s(k)/eta);     % count no. cells
    dc = c(k) - c(k-1);         % calculate cell growth
    if dc>0                     % create the new cells and choose their delays
        tau = exp(sigma*randn(dc,1)+mu);
        gr(c(k-1)+1:c(k),2) = tau;
    end;

end;

% --- done generating events; extract results
[dt k] = deal(diff(fk(:,2)),(1:size(fk,1)-1));

% --- plot resulting development curve
figure(1); clf;
loglog(k,dt,'r-','LineWidth',2); hold on;
loglog([1 nok],exp(mu).*([1 nok]).^(-1),'k--','LineWidth',3); hold off;
xlabel('Cumulative number of events, \it{k}','FontSize',16);
ylabel('Time to next event, \Delta\it{t} \rm{(days)}','FontSize',16);
set(gca,'FontSize',16,'YTick',10.^(-6:4));
h1=legend(strcat('Simulation, \nu/\eta=',num2str(nu/eta,'%3.1f')), ...
     'Model, \Deltat\propto k^-^1',1); set(h1,'FontSize',16);
\end{verbatim}
\end{footnotesize}
\baselineskip14pt

\section{Statistical Model for the Frequency of Attacks}
\label{appendix:C}
The probabilistic model for event delays used in the main text, given by Eq.~(1), has the precise form of
\begin{align}
\Pr(\deltat \,|\, k) & = \left(\frac{\sqrt{2/\pi}}{ \sigma \left(1 - {\rm Erf}\!\left[\frac{\beta \log k-\mu}{\sigma\sqrt{2}}\right]\right)}\right)   \,{\rm exp}\!\left[\frac{-(\log \deltat+\beta\log k-\mu)^{2}}{2\sigma^{2} }\right]  \label{eq:model:full}
\end{align}
where the leading term is the normalization constant and ${\rm Erf}(\cdot)$ is the error function. In words, this model asserts that the logarithm of the delay $\deltat$ is a random variable distributed according to a Normal distribution $\mathcal{N}(\nu,\omega)$ (or equivalently, the delay is log-normally distributed) with a lower cutoff at $\deltat=1$ day (to reflect the timing resolution of the event data), constant variance $\omega$ and a distributional mean $\nu$ that decreases systematically with increasing experience $k$.
In Eq.~\eqref{eq:model:full}, the parameter $\mu$ denotes the characteristic delay between attacks, and in particular the delay between the first and second attacks, while $\sigma^{2}$ denotes the variance in the expected delay.

The equation given in the main text for the expected delay as a function of experience---the central tendency of the conditional distribution of delay as a function of experience---can be derived in the usual way. Doing so yields
\begin{align}
{\rm E}[\log \deltat] = \mu - \beta \log k + \left(\frac{ {\rm exp}\!\left[ \frac{-(\beta \log k - \mu)^{2}}{2\sigma^{2}}  \right] \sqrt{2/\pi}   }{\sigma^{-1}\left(1-{\rm Erf}\!\left[ \frac{\beta \log k - \mu}{\sigma\sqrt{2}} \right]\right)} \right)\enspace ,
\label{eq:expected:delay}
\end{align}
which has a simple leading form and a complicated trailing term. For small values of $k$, the expected delay is dominated by the leading two terms, i.e., the trailing term is small in relative magnitude, and thus the trend is well-approximated by a power-law function $\deltat\approx \rm{e}^{\mu} k^{-\beta}$, where $\rm{e}^{\mu}$ represents the initial rate of attack of a group. At larger values of $k$, the expected delay is dominated by the trailing term, which makes the expected delay to approach $\deltat=1$ more slowly than a power law.%

When fitting this model to the empirical data, we estimate its parameters using standard numerical procedures to maximize the likelihood of the data (in this case, the Nelder-Mead 1965~\nocite{nelder:mead:1965} method). Standard error estimates for the uncertainty in the parameters are then estimated using a bootstrap procedure on the organizations in the sample.

The striking ``data collapse'' shown in Figure~3b illustrates that the conditional probability distributions do indeed align closely with the estimated log-normal model for delays. Why delays should be log-normally distributed remains a mystery.

Finally, we point out that very few groups (e.g., Hamas, Fatah, LTTE, FARC, etc.) manage to become highly experienced ($k\gtrsim100$). This means that the fit of the model for large-$k$ is primarily controlled by the delays at much smaller values of $k$, where the vast majority of the data lay. This fact explains the slight misfit of the model to the delays for highly experienced groups. However, it also highlights the fact that the behavior of inexperienced groups early in their lifetime is highly predictive of the behavior of mature organizations.

\section{Domestic vs.\ Transnational Events}
\label{appendix:D}
From 1968--1997, the MIPT event database was maintained by RAND as part of its project on transnational terrorism. As a result, almost no domestic terrorist attacks are included before 1998, after which the scope of the database was significantly expanded (in part due to the Oklahoma City bombing in 1995) to include purely domestic events worldwide. Although organizations and events are not coded as being transnational or domestic, the inconsistency in database scope provides an opportunity to test whether the frequency dynamics of domestic terrorism organizations differs from those of transnational organizations.

By dividing events into those generated by organizations whose first event occurred 1968--1997 and those generated by organizations whose first event occurred in 1998--2008, and then repeating the frequency-curve analysis from the main text, we may test whether the frequency-acceleration phenomena appears only in one time period or the other. Further, because events in the 1998-2008 period are mainly domestic events, while those in the 1968--1997 period are only transnational events, the two time periods serve as proxies for transnational-only and domestic-only terrorism. This division does not control for non-stationary effects.

\begin{figure}[t!]
\begin{center}
\includegraphics[scale=0.5]{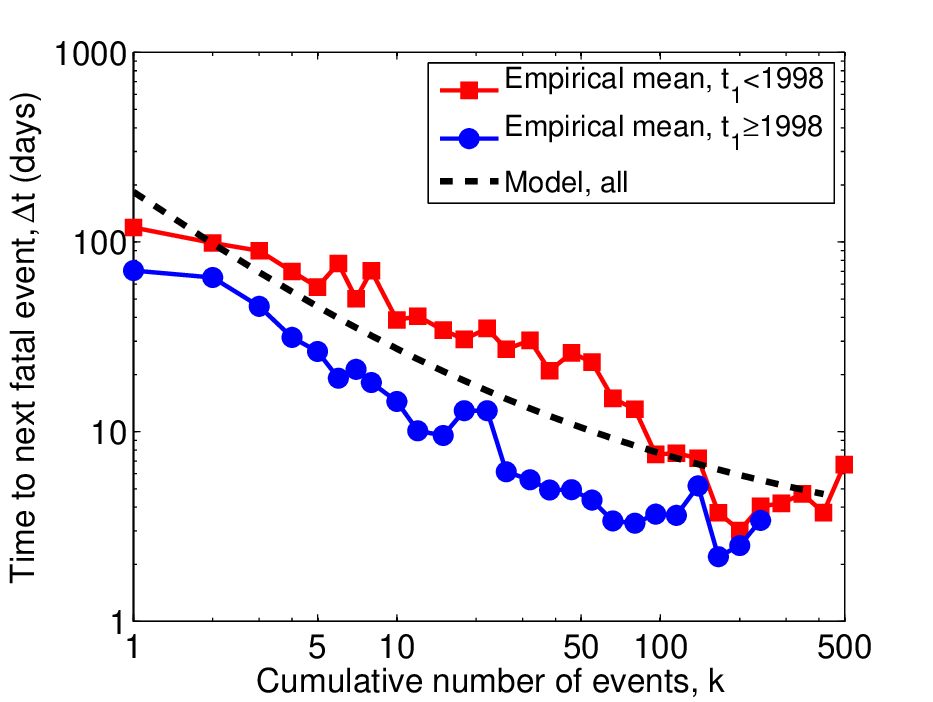}
\end{center}
\vspace{-3mm} \caption{The attack frequency development curves, plotted as the average delay versus experience, for groups whose first attack was in 1968--1997 versus those whose first attack was in 1998--2008, along with the model estimated for all events from the main text.}
\label{fig:1998}
\end{figure}

Figure~\ref{fig:1998} shows that the development curve phenomenon
is robust to this division, indicating that the
frequency-acceleration appears to hold for both transnational and
domestic terrorism. One difference between these time periods does
emerge: the rate of acceleration for the 1968--1997 data
(transnational only) is $\hat{\beta}_{t_{1}\leq1997}=1.0\pm0.2$
(stderr), statistically indistinguishable from the analysis of all
organizations in the main text, while the estimated acceleration
for the 1998--2008 data (mainly domestic) is slightly faster, with
$\hat{\beta}_{t_{1}>1997}=1.3\pm0.2$. The origin of this
difference may be related to the increasing frequency of
religiously-motivated terrorism in the 1990s and
beyond~\cite{rapoport:2004,rosenfeld:2011}, who collectively
exhibit a lower value of $\hat{\beta}$ than other types of
terrorism. An interesting alternative explanation, however, is
that some non-stationary process is having a consistent upward
pressure on $\beta$ over time, for all organizations. One
candidate process is the development and spread of modern
communications and digital technology, which may enable more
widespread or effective recruiting efforts and thus faster
organizational growth.

\section{Political Ideology \& Frequency and Severity Curves}
\label{appendix:E}
Our results for the developmental dynamics of event frequency and severity are good descriptions of the generic behavior of terrorist organizations. However, we have so far omitted any role for organizational covariates, many of which are believed to have important impacts on organizational behavior and decisions (see~\cite{asal:rethemeyer:2008,pape:2003,clauset:etal:2010:a}, among others). We investigate this question by studying the impact, if any, political or ideological motivation may have on the frequency curve's structure; we leave the investigation of other covariates for future work.

\begin{figure}[t]
\begin{center}
\includegraphics[scale=0.435]{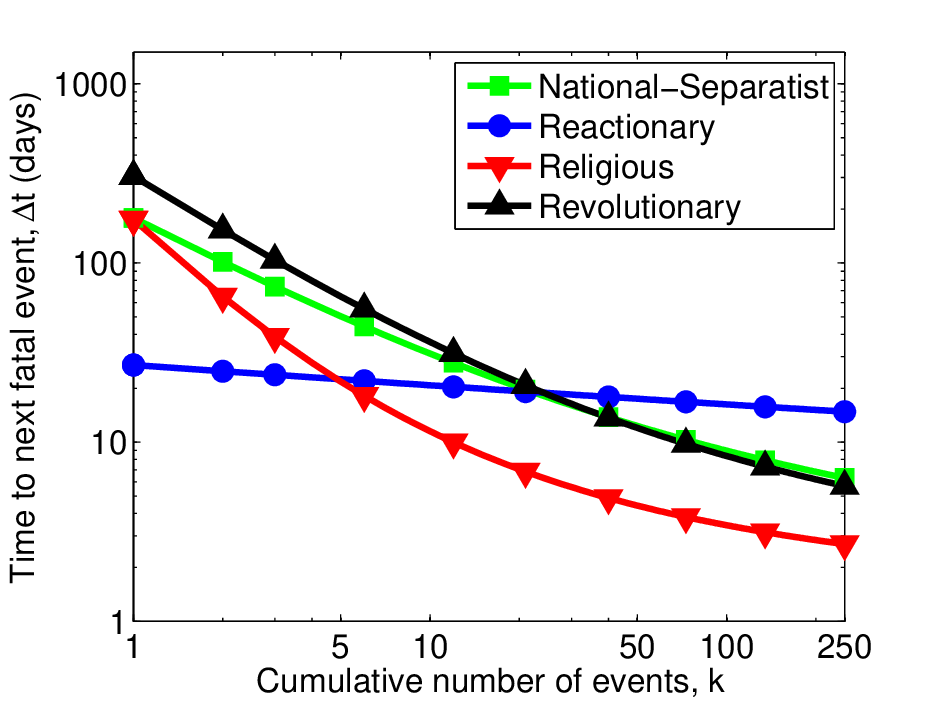}
\includegraphics[scale=0.435]{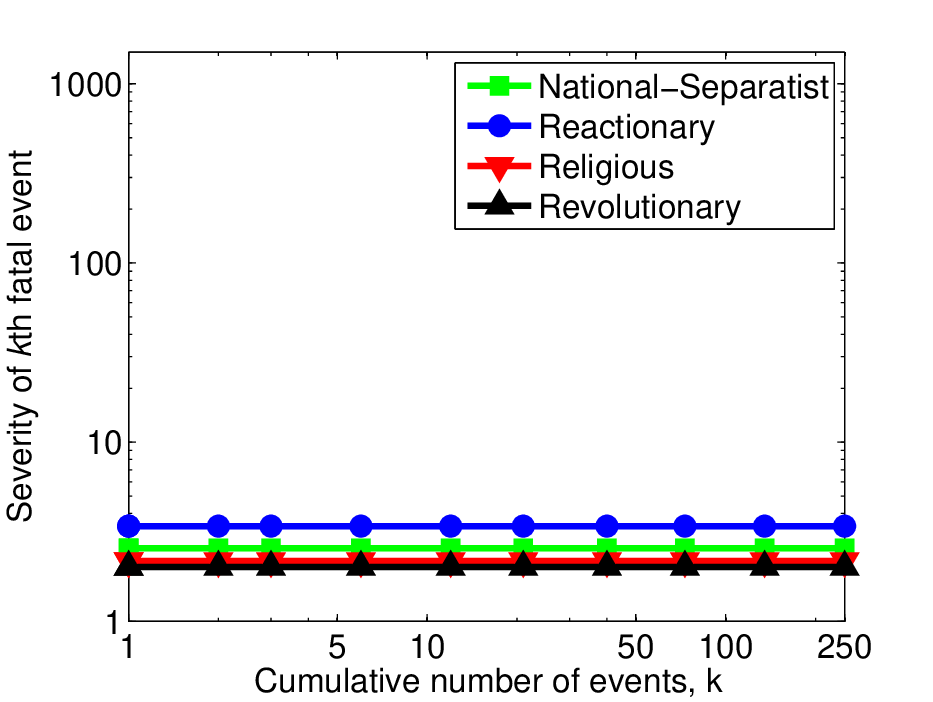}
\end{center}
\vspace{-3mm}  \caption{(a) Estimated frequency curves for four ideological categories, showing that religious groups develop extremely quickly relative to other types. (b) Estimated severity curves for the same categories, showing the same pattern of independence as Fig.~4a.}
\label{fig:ideology:curves}
\end{figure}

Miller~\cite{miller:2007} divides the political motivations for terrorism or group ideologies into four conventional categories: nationalist-separatist, reactionary, religious and revolutionary. We coded according to Miller's criteria the 131 groups in our sample with $k\geq10$ deadly events, who together account for 85\% of events (the majority of our data), and fitted Eq.~(1) to the data within each ideological category. Organizations with multiple political motivations were placed in multiple categories, which would only lessen any differences between estimated parameters for different categories. Fig.~\ref{fig:ideology:curves}a shows the corresponding central tendencies, as described by Eq.~(2). Table~3 summarizes the estimated parameters for each ideological category and groups overall.

We again test the statistical significance of the acceleration effect within each ideological model using a two-tail test against a null model with fixed $\beta=0$ (no acceleration over time). In all cases, the estimated $\beta$ parameter is highly statistically significant (at the $p<0.001$ level), indicating that the acceleration within each category is real.

Among the four ideological categories, we observe wide variation in the estimated values of $\beta$ and thus in the strength of the feedback loop governing the frequency of attacks. Religious groups have the largest value at $\hat{\beta}=1.7\pm0.5$, placing them firmly in the super-linear feedback regime and implying very strong acceleration in the frequency of attacks over time. In contrast reactionary organizations have the smallest at $\hat{\beta}=0.1\pm0.3$, placing them strongly in the sub-linear regime. Revolutionary and nationalist-separatist categories are statistically indistinguishable from the linear-feedback regime of $\beta=1$.

The typical religious group, i.e., one accelerating along the generic production trajectory identified above, with $k=10$ deadly attacks, attacks as frequently as the typical revolutionary group with $k=51$ deadly attacks or the typical nationalist-separatist group with $k=129$ attacks. When viewed in terms of calendar time, this difference is even more striking: it takes the typical religious terrorist organization only 400 days (1.1 years) to generate its first 10 attacks and at this point its production rate is approximately one attack every 5 days. In contrast, the typical revolutionary organization takes 1666 days (4.6 years), more than four times as long, and a typical nationalist-separatist organization takes 2103 days (5.8 years), to achieve an equal production rate. Combining this insight with the results of our static analysis on the role of size, the explosive acceleration by religious groups implies that they grow in size extremely quickly, which is the ultimate cause of their dramatic production rates.

But religious organizations are not universally more dangerous. Comparing the $\hat{\mu}$ parameters, which governs the characteristic delay between subsequent attacks, we observe a more complicated story: reactionary groups initially attack the fastest, with the fitted model estimating typically $\deltat=47$ days between their first and second attacks, while all other groups take substantially longer ($\deltat>100$ days). This difference in initial production rates is quickly eliminated by the explosive acceleration of religious groups as well as the more measured development of revolutionary and nationalist-separatist organizations, whose typical event production rates overtake that of reactionary groups after between 5 and 25 events.

Much previous work on religious terrorism has argued, largely on theoretical grounds, that such organizations are fundamentally more dangerous than secular groups~\cite{miller:2007,rapoport:1984,hoffman:1995,juergensmeyer:1997} because they have fewer social restrictions on their activities and are thus more free to produce and target violence than secular organizations, whose victims may be potential sympathizers. Our results provide indirect support for this argument, in the sense that religious organizations exhibit explosive acceleration in the production of violence while secular organizations exhibit more moderate acceleration.

However, arguments that religious organizations are universally more dangerous may have over-simplified organizational behavior by ignoring how organizations may change their behavior over time and how they vary relative to other organizational types. We find that very early in their life histories, religious groups are in fact less dangerous than reactionary groups, and only slightly more dangerous than national-separatist or revolutionary groups. It is only over the long term that the explosive acceleration experienced by religiously-motivated organizations allows them to cumulatively produce so many more events than other types of organizations. That is, only if a religious organization succeeds in reaching a more mature state does it pose a greater overall risk than groups with secular motivations. And, it is important to note that historically speaking, most organizations do not live so long~\cite{cronin:2009}: fully 55\% of organizations in the MIPT database are associated with only a single event.

\begin{table}
\caption{Severity curve parameters for organizations with similar political motivations. Note: statistical significance calculated using a $t$-test on Pearson's correlation coefficient.}
\vspace{2mm}
\centering
\begin{tabular}{r|cc|cr|c}
political motivation & groups & events & $\langle x \rangle$ & $r$ & significance \\
\hline
nationalist-separatist & 51 & 1003 & $6.1$ & $0.0071$ & $p=0.75$\\
reactionary & 5 & 77 & $7.1$ & $0.1194$ & $p=0.27$\\
religious & 17 & 753 & $5.2$ & $-0.0062$ & $p=0.49$\\
revolutionary & 41 & 725 & $5.1$ & $-0.0109$ & $p=0.38$\\
\hline all groups & 381 & 3143 & $7.3$ & $-0.0240$ & $p=0.17$
\end{tabular}
\end{table}

Turning briefly to the question of how event severity varies with organizational ideology, we repeat the same severity-curve analysis on the deadly events produced by the 131 highly prolific organizations. Figure~\ref{fig:ideology:curves}b shows the resulting ideology-specific severity curves and Table~4 summarizes the estimated model parameters, where the model now is a simple linear regression of severity~$x$ against experience~$k$. As above, we find no systematic dependence of severity of attacks on organizational experience within any of the ideological categories. That is, none of the model coefficients are significant, and the average severity of events within each category vary only a little. Thus, we find that political ideology has no systematic impact on the severity of events or the trajectory that event severities take over the lifespan of an organization.

\end{appendix}

\end{document}